\documentclass[a4paper,12pt]{article}
\usepackage[cmex10]{amsmath}
\usepackage{float}
\usepackage{graphicx}
\usepackage{array}
\usepackage{color}
\usepackage{fixltx2e}
\usepackage{authblk}
\usepackage{rotating}

\begin{document}

\title{Design and Characterization of Q-enhanced Silicon Nitride Racetrack Micro-Resonators}

\author[1]{Pedro~Chamorro-Posada}
\author[2]{Roc\'{\i}o~Ba\~nos}
\affil[1]{Dpto. de Teor\'ia de la Se\~{n}al y Comunicaciones, 
  Ingenier\'ia Telem\'atica,
  Universidad de Valladolid, Paseo Bel\'en 15, 47011 Valladolid, Spain.
  (e-mail: pedcha@tel.uva.es)}
\affil[2]{VLC Photonics S.L., Camino de Vera
s/n, 46022 Valencia, Spain}

\maketitle
\begin{abstract}

 Q-enhanced racetrack micro-resonators for the silicon nitride photonics integration platform have been designed, fabricated and characterized.  The proposed geometries  permit to mitigate the impact of radiation loss at curved waveguides, one of the major limitations of silicon nitride circuits, therefore providing an increase of the intrinsic Q factor of micro-resonators when compared with the conventional structures with the same bent radii.  The schemes put forward in this work permit a reduction of the size of the devices that has a direct impact on the integration scale in this platform.   When used in the curved sections of waveguides routing optical signals within an integrated photonic circuit, these geometries provide a reduction of the radiation loss and permit the use of smaller bent radii and to increase the circuit integration density.

\end{abstract}

\section{Introduction}

Racetrack microresonators are key building blocks in the different photonic integration platforms \cite{bogaerts}.   They permit to implement add-drop multiplexers \cite{little1}, optical filters \cite{little2,chamorro11}, optical switches \cite{almeida}, sensors \cite{vos}, modulators \cite{xu}, and they are the constituent elements of slow and fast light systems based on coupled resonator waveguides \cite{scheuer,chamorro09}.  

Irrespective of the particular application, the behavior of a micro-resonator in a optical system is defined by its Q-factor, which is built from two distinct contributions:  one is the coupling to the external circuit, which is design-specific, and the other is due to the internal loss that define the unloaded or intrinsic Q-factor.  In general, high unloaded Q-factor or, equivalently, low loss is a desirable property in any implementation.
 
Propagation loss in the resonator can have various origins:  the intrinsic loss of the waveguide material, the effect of the roughness of the waveguide walls or the curving of waveguide sections.  The impact of each loss mechanism depends on the specific platform.  In silicon on insulator (SOI) photonic circuits, the effect of bending can be negligible even for very small radii of curvature because of the highly confining high index contrast silicon waveguides.  Nevertheless, intrinsic propagation losses are relatively high in the SOI platform.  The converse situation is found in Si\textsubscript{3}N\textsubscript{4}/SiO\textsubscript{2} photonic integrated circuits, with very small intrinsic losses and large radiation losses due to waveguide curvatures, for the same bent radii, due to the reduced waveguide core/cladding refractive index contrast.

In \cite{chamorro}, a general approach for the Q enhancement of racetrack micro-resonators was proposed.  This strategy acts on the two sources of optical loss due to waveguide bending in a racetrack micro-resonator.  In the curved waveguide sections, the modal effective indices become complex \cite{hiremath} and field propagation is accompanied by the continuous production of a radiation wave.  The second cause of loss is due to the discontinuities existing at the straight-bent sections that produce localized radiation beamed along the optical axis \cite{chamorro}. In the scheme of \cite{chamorro}, the first effect is reduced with the aid of bent coupled asymmetric waveguide sections.  This geometry modification is compatible with the conventional lateral offset technique \cite{kitoh} that is introduced at the discontinuities to mitigate the second source of radiation.  These two combined counteractants permit to notably reduce the losses and enhance the Q-factor of racetrack microresonators in low-contrast integration platforms like silicon nitride, silica on silicon or polymer.  In turn, this Q-enhancement allows the use of shorter bend radii, with a direct impact on the attainable integration density.  

The 2D numerical calculations presented in \cite{chamorro} are adequate for a proof of concept analysis but lack the high accuracy required in the design of devices targeted for the fabrication of integrated circuits.  In \cite{radiacion}, a full vector 3D extension of the previous work was presented including the application of modal calculations \cite{wgms3d,krause} that hold the high level of accuracy demanded by this problem.  In this work, we present the results of the design and characterization of the racetrack geometries proposed in \cite{chamorro} fabricated in the silicon nitride platform.  Radiation quenching structures based on bent coupled waveguides are experimentally realized for the first time in this work.

\begin{figure}[!t]
  \centering
  \begin{tabular}{c}
    \includegraphics[width=.4\columnwidth]{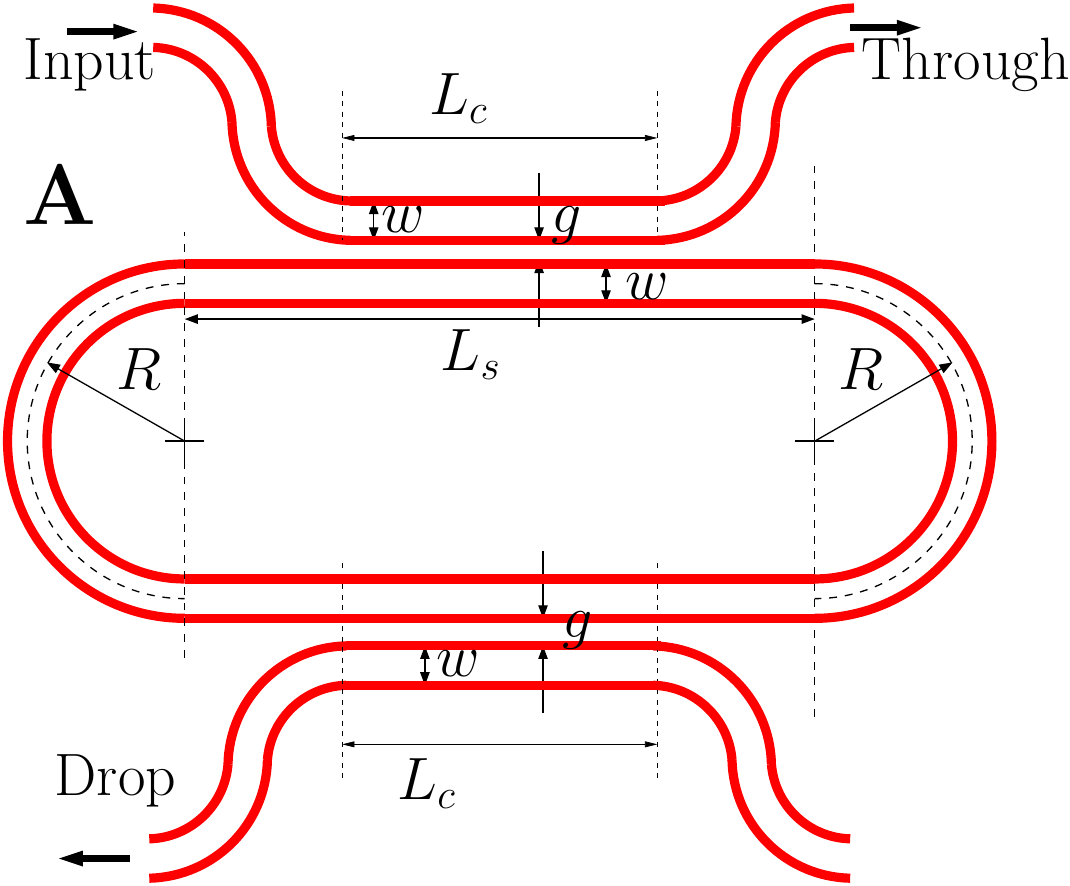}\\
    \includegraphics[width=.45\columnwidth]{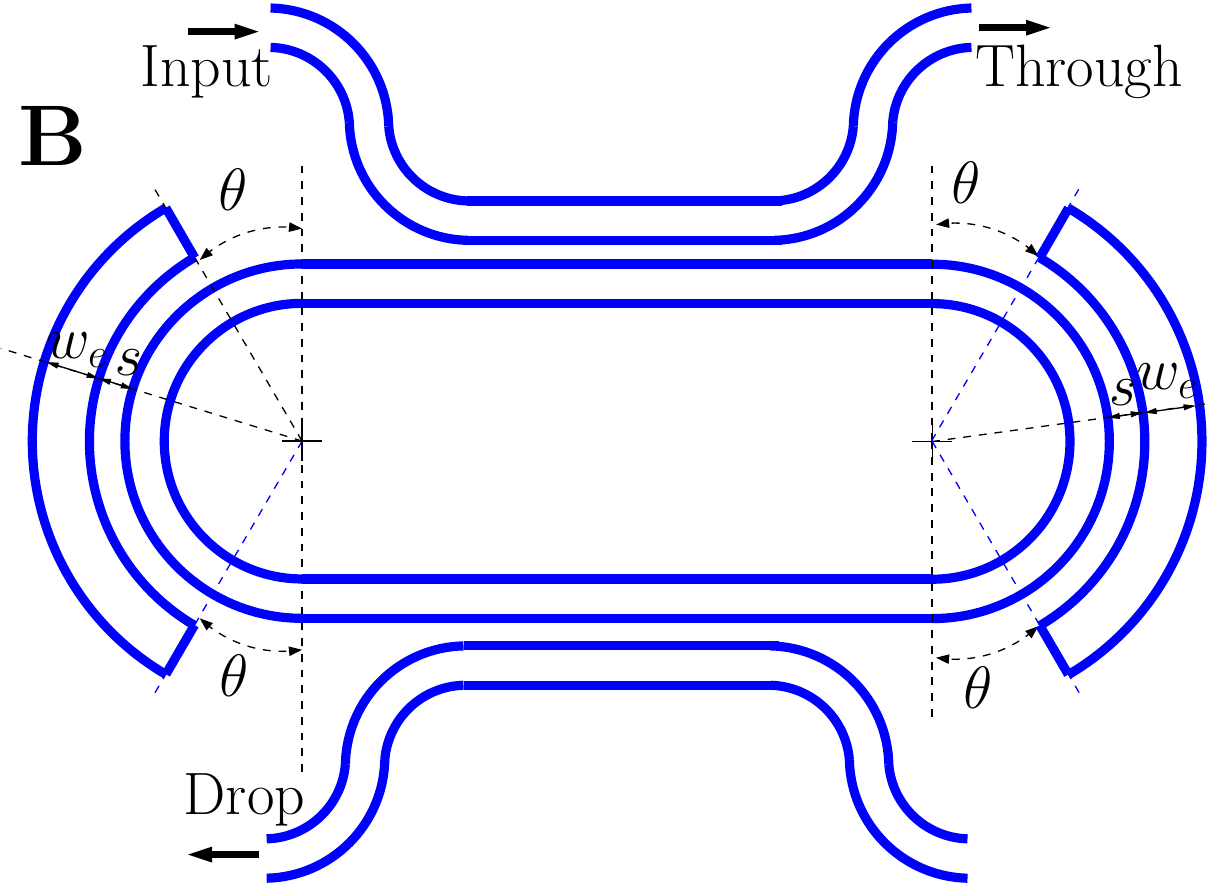}\\
    \includegraphics[width=.45\columnwidth]{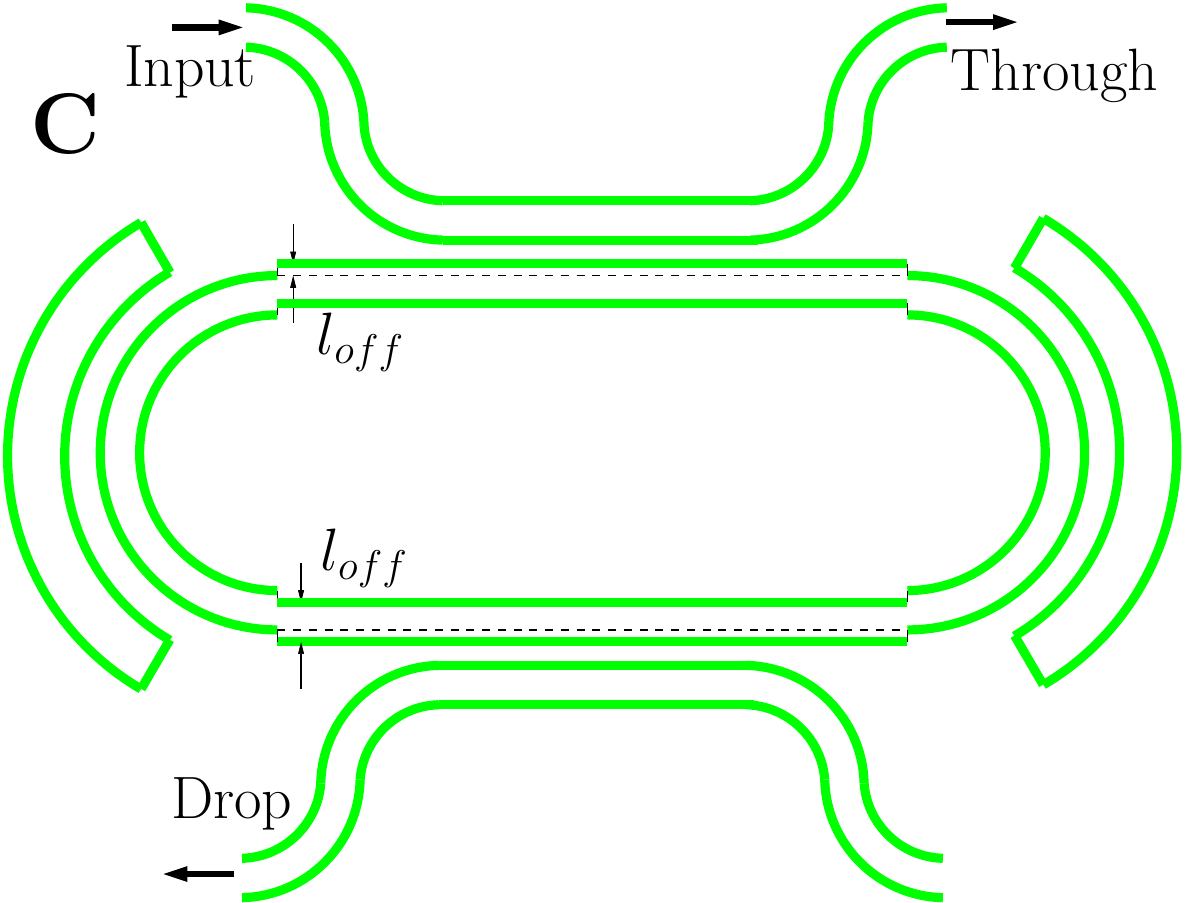}
  \end{tabular}
\caption{A is the geometry of a conventional racetrack micro-resonator in an add-drop configuration. Case B incorporates radiation quenching coupled asymmetric curved waveguides and C also includes lateral offsets at the straight-bent discontinuities.}\label{fig::estructuras}
\end{figure}

\begin{figure}[!ht]
\centering
\begin{tabular}{c}
  {\large (a)}\\
  \includegraphics[width=0.4\columnwidth]{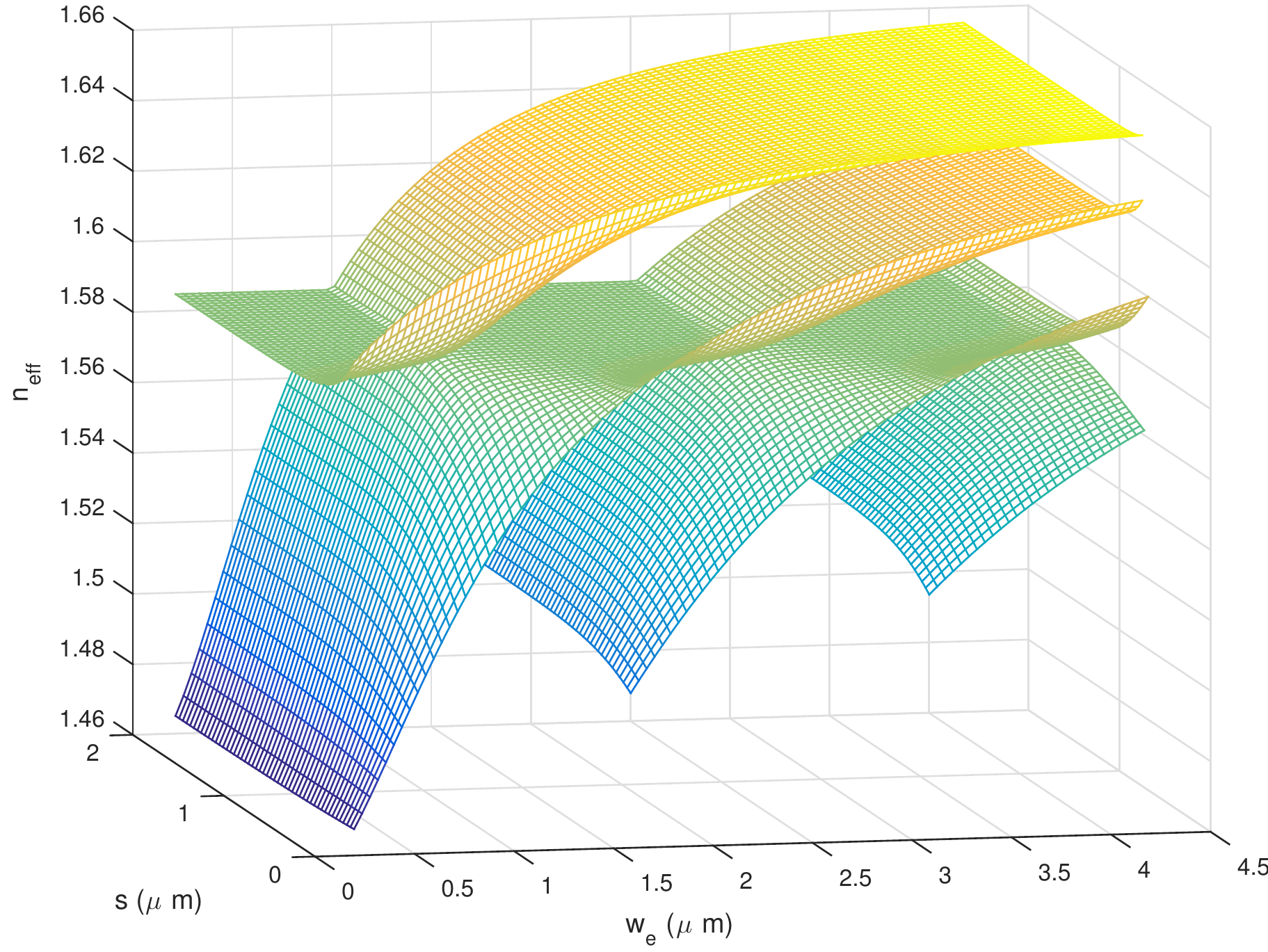}\\
  {\large (b)}\\
  \includegraphics[width=0.4\columnwidth]{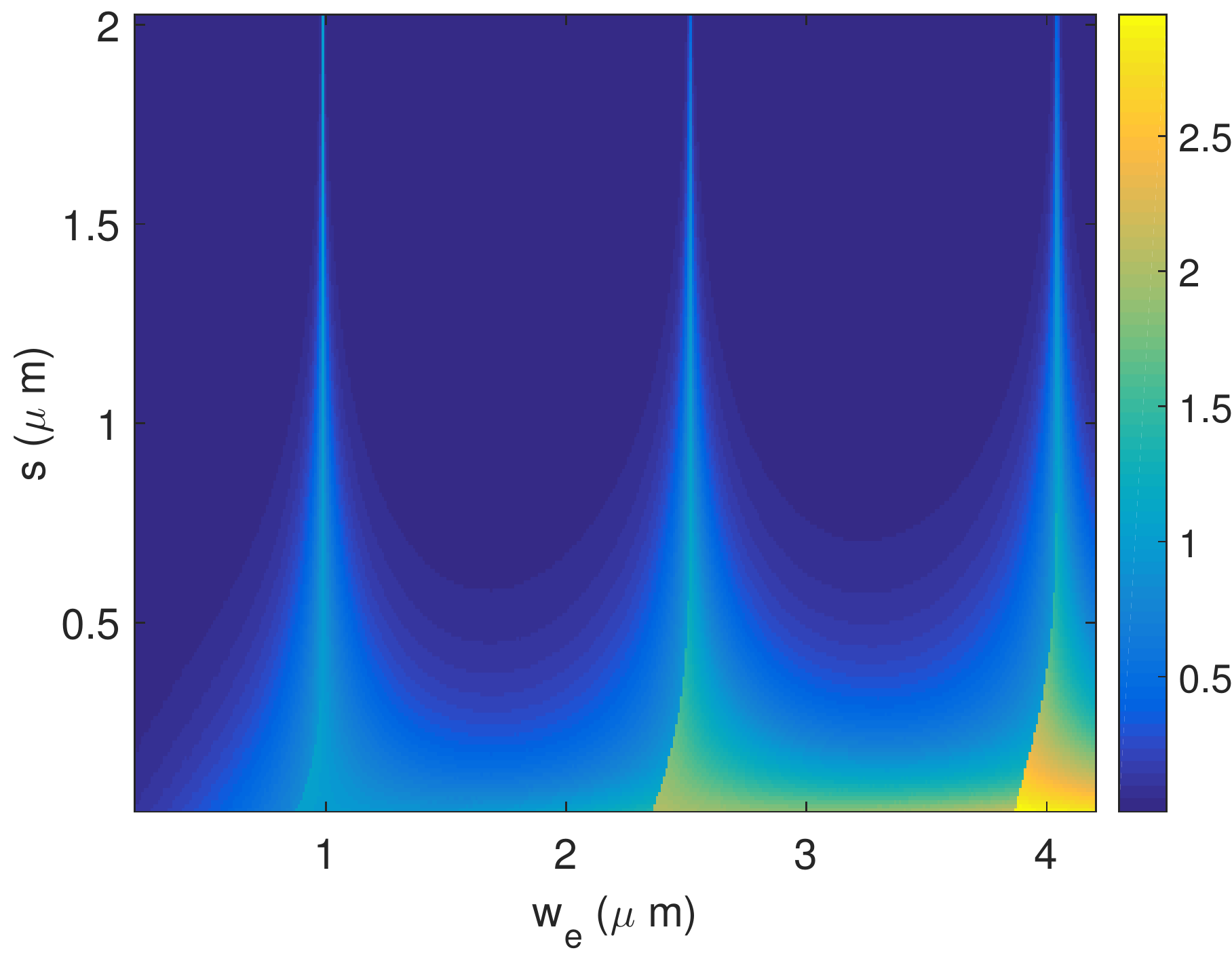}
  
\end{tabular}
\caption{(a) Modal effective indices of quasi-TE modes of a straight asymmetric coupler as a function of $w_e$ and $s$. (b) Ratio of the power carried by the core of the rightmost waveguide $P_2$ to that carried by the core in the main waveguide $P_1$ for the mode predominantly guided in the main waveguide as a function of $w_e$ and $s$. Calculations have been performed using the effective index method.}\label{fig::EIM}
\end{figure}

\begin{figure}[!ht]
\centering
\begin{tabular}{c}
  {\large (a)}\\
  \includegraphics[width=0.4\columnwidth]{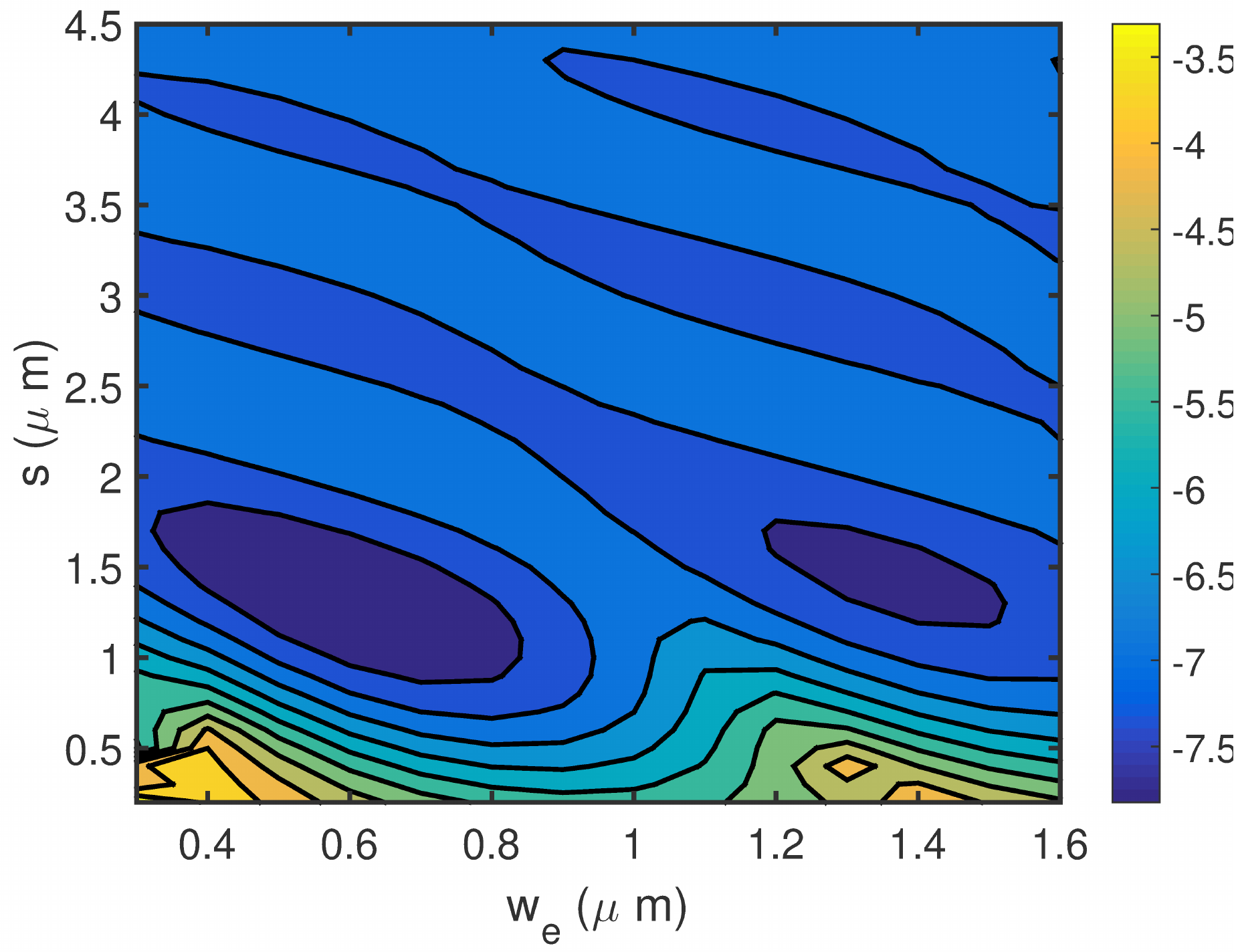}\\
  {\large (b)}\\
  \includegraphics[width=0.4\columnwidth]{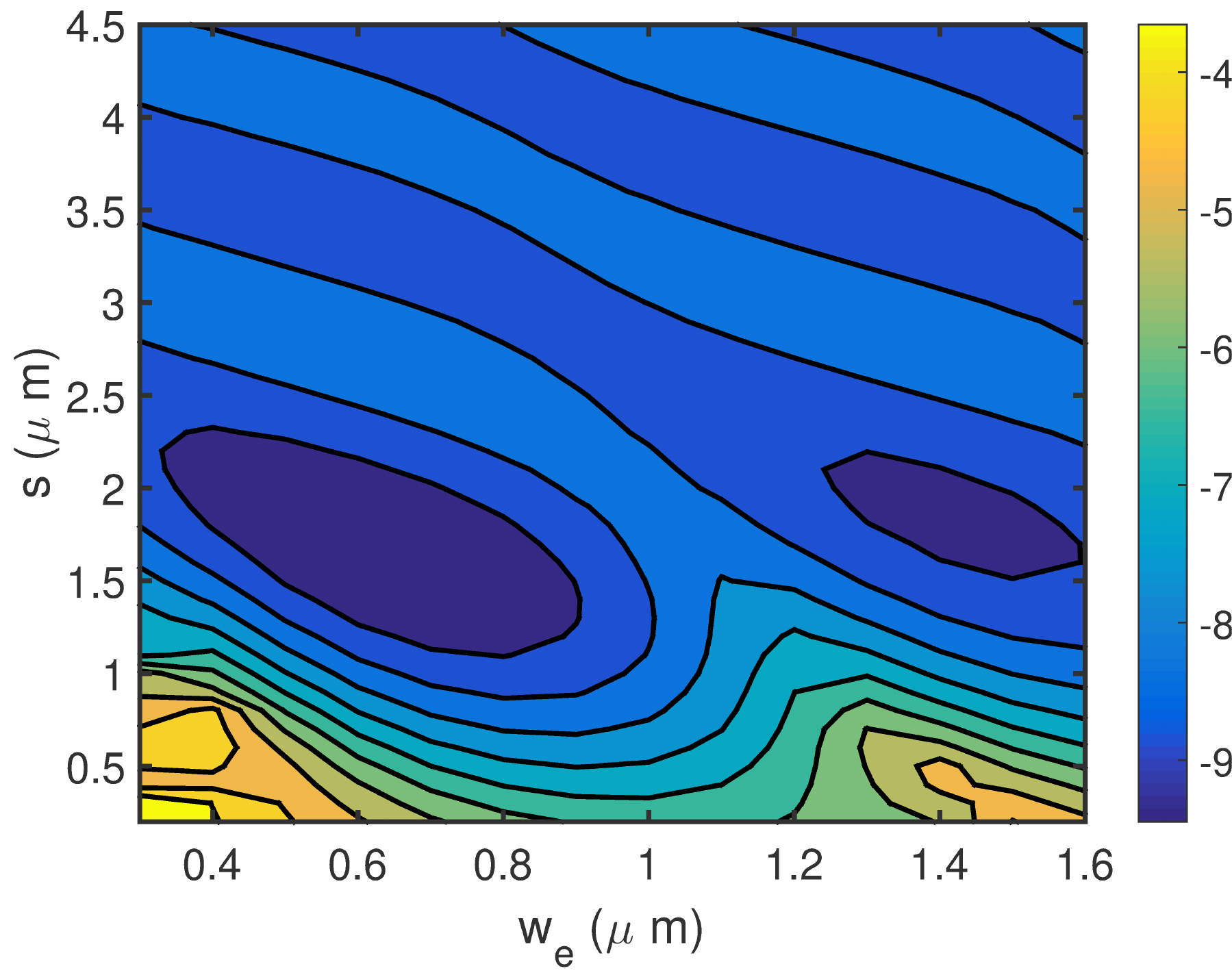}\\
  {\large (c)}\\
  \includegraphics[width=0.4\columnwidth]{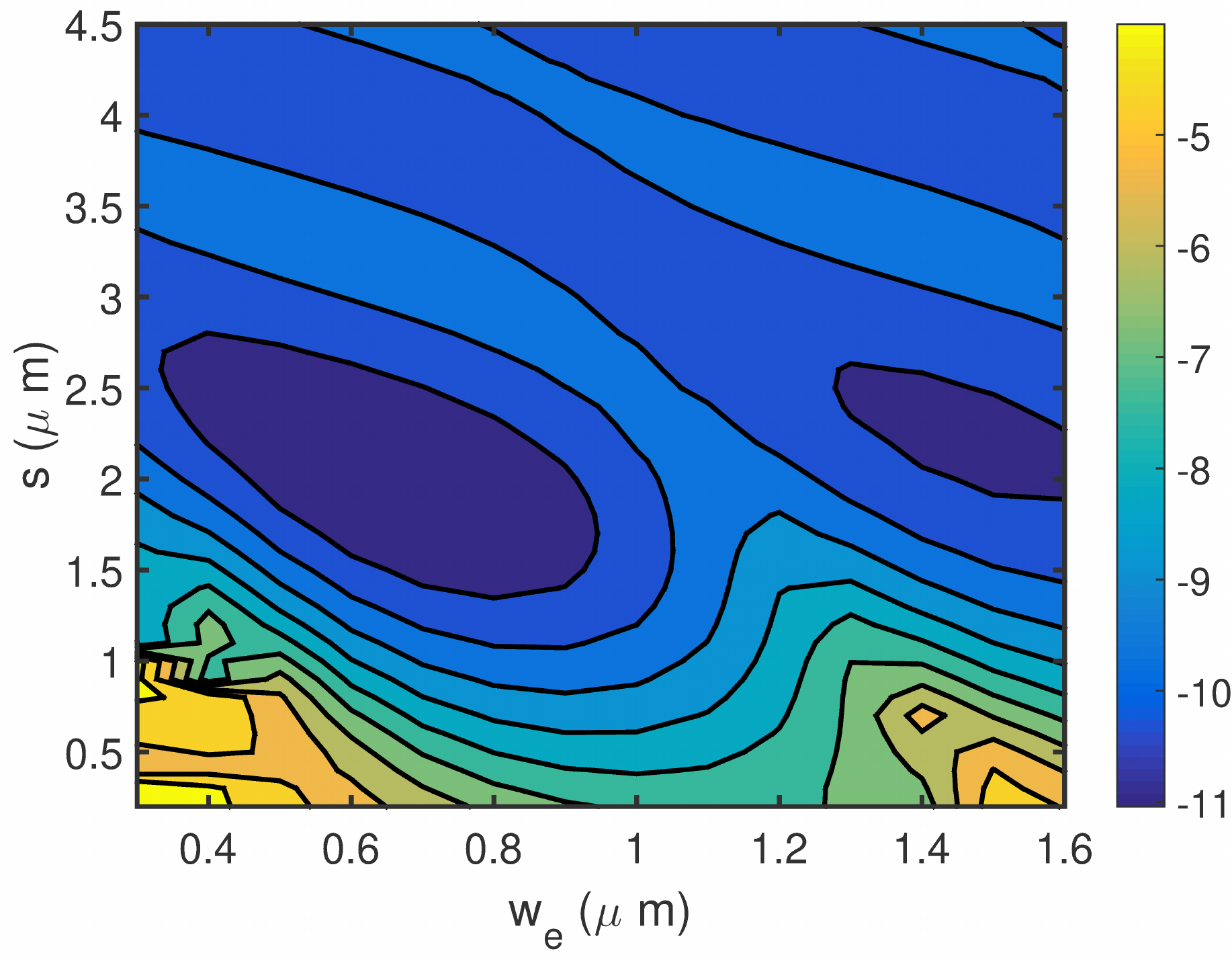}
\end{tabular}
\caption{Contours of $log_{10}\left(n_i\right)$ for the quasi-mode localized at the interior waveguide.  $R=15$ $\mu$m, $R=20$ $\mu$m, and $R=25$ $\mu$m for (a), (b), and (c) plots, respectively.}\label{fig::contornos}
\end{figure}

\section{Q-enhanced racetrack microresonators}

  The geometries of the three fabricated structures are displayed in Fig. \ref{fig::estructuras}.  Subplot A, shows a conventional racetrack geometry in an add-drop configuration, the geometry depicted in subplot B includes radiation-quenching curved asymmetric coupled sections and the design in C implements the former plus lateral offsets at the discontinuities.

In the schematics of Fig. \ref{fig::estructuras}, $w$ is the waveguide width, $L_s$ is the length of the straight sections of the racetrack micro-resonator, and $R$ is the bend radius of the curved sections.  The total ring length is given by $L=2(\pi R+L_s)$.  In all the fabricated structures $L_s=72$ $\mu$m.  The self-coupling coefficients of the evanescent couplers that connect the resonator to the external waveguides, which we will assume to be equal and of value $r$, is determined by the effective coupler length, which depends on both $L_c$ and the curvature of the access waveguides \cite{xia}, and the gap $g$ between the ring and access straight waveguides.  Symmetric structures, with identical couplers in the upper and lower branches are assumed.

In an uniformly curved waveguide geometry of radius $R$, the quasi-guided mode solutions can be described as \cite{radiacion}

\begin{equation}
\left\lbrace\mathbf{E},\mathbf{H}\right\rbrace (\rho,\phi,z)=\left\lbrace\mathbf{e},\mathbf{h}\right\rbrace (\rho,z)\exp\left(j\beta R \phi\right),\label{modo}
\end{equation}
where $(\rho,\phi,z)$ define a cylindrical coordinate system having $\rho$ and $z$ as the coordinates transverse to the propagation \cite{radiacion}.  $\mathbf{E}$ and $\mathbf{H}$ are, respectively, the electric and magnetic field strengths,  $\left\lbrace\mathbf{e},\mathbf{h}\right\rbrace (\rho,z)$ are the complex modal fields, $R$ is the bend radius, and $\beta$ is the complex propagation constant.  The complex effective index of the modal field $n=n_r+jn_i$ is related to the propagation constant as $\beta=k(n_r+jn_i)$ with $k=\omega/c$.  These modal solutions continuously shed radiation as they propagate and the imaginary part of the complex effective index $n_i$ accounts for the radiation loss \cite{hiremath}.  

\begin{figure}[!t]
\centering
\begin{tabular}{c}
  {\large (a)}\\
  \includegraphics[width=0.4\columnwidth]{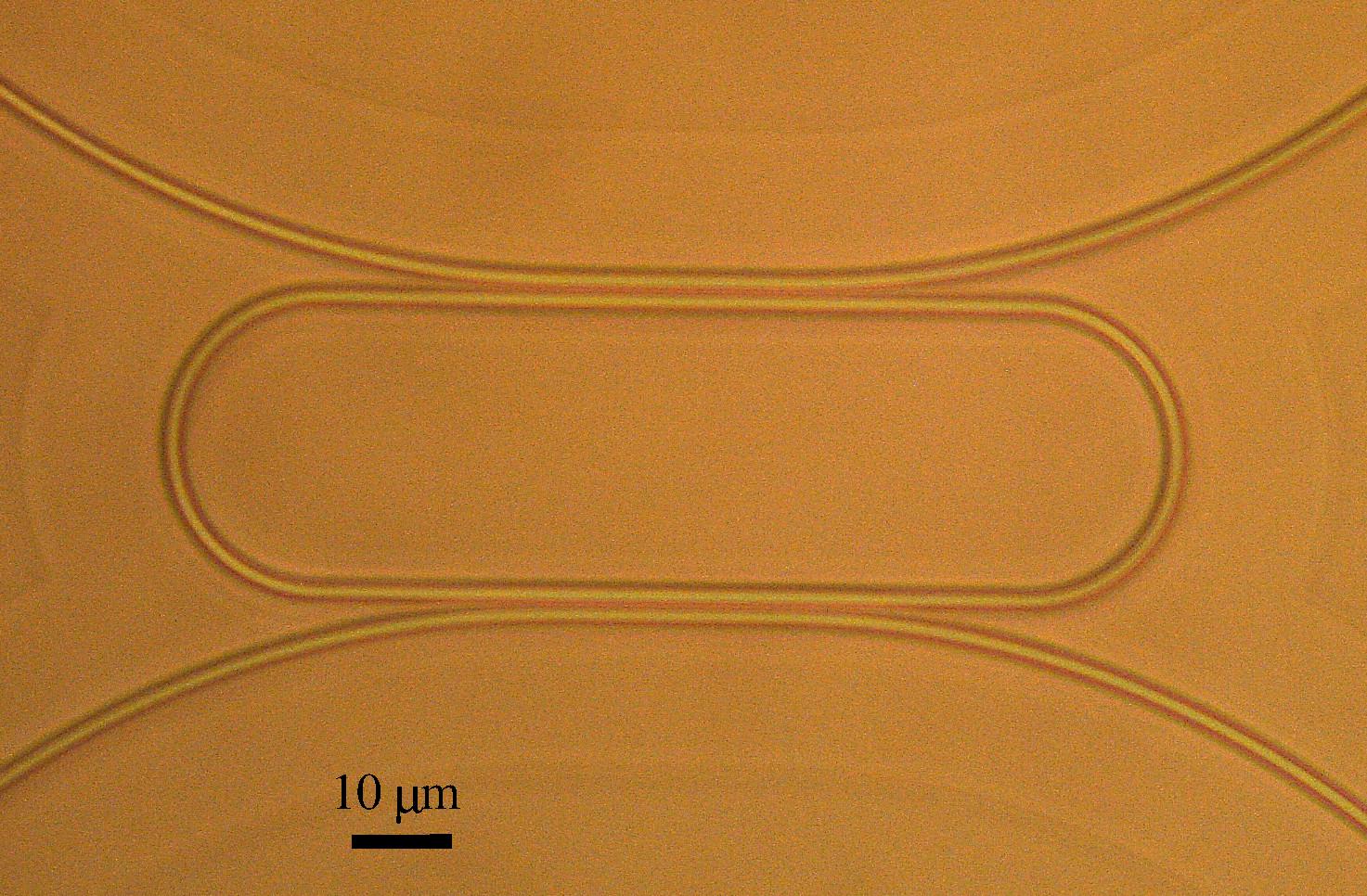}\\
  \\
  {\large (b)}\\
\includegraphics[width=0.4\columnwidth]{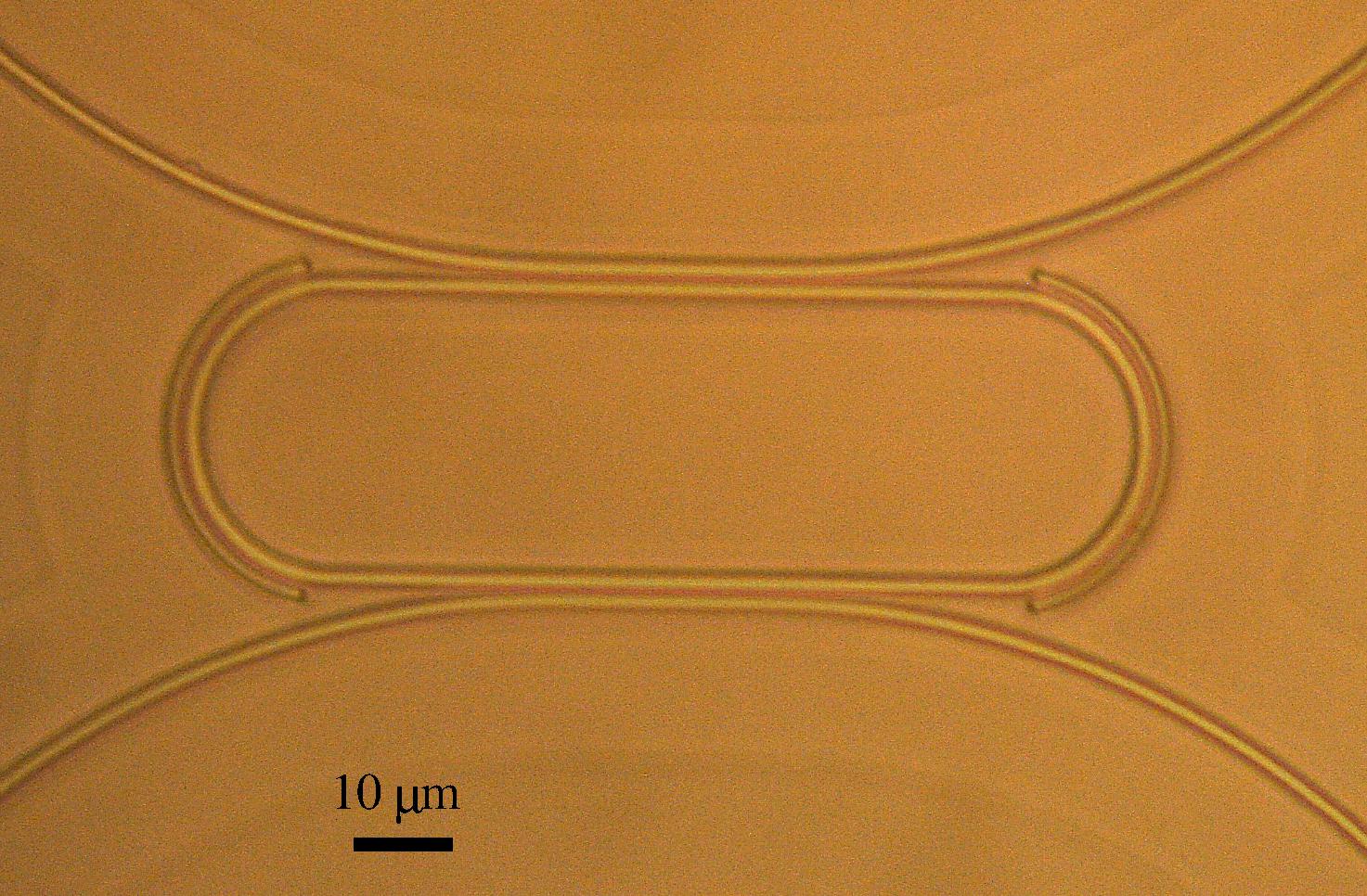}
\end{tabular}
\caption{Micorgraphs of the fabricated micro-resonators of type A and B (subfigures (a) and (b), respectively) for $R=15$ $\mu$ m.}\label{fig::imagenes}
\end{figure}

\begin{figure}[!t]
\centering
\begin{tabular}{c}
  {\large (a)}\\
  \includegraphics[width=0.4\columnwidth]{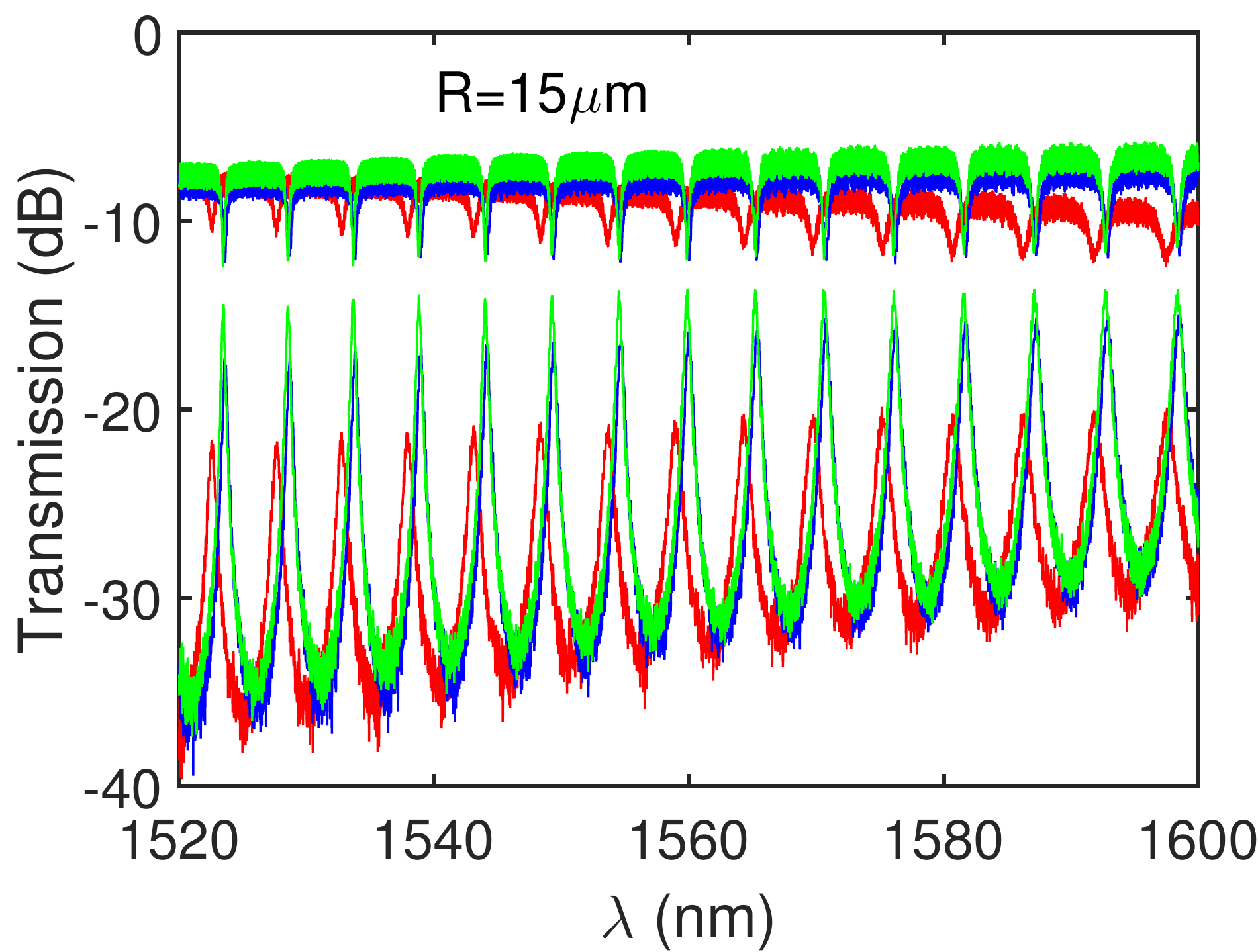}\\
  {\large (b)}\\
  \includegraphics[width=0.4\columnwidth]{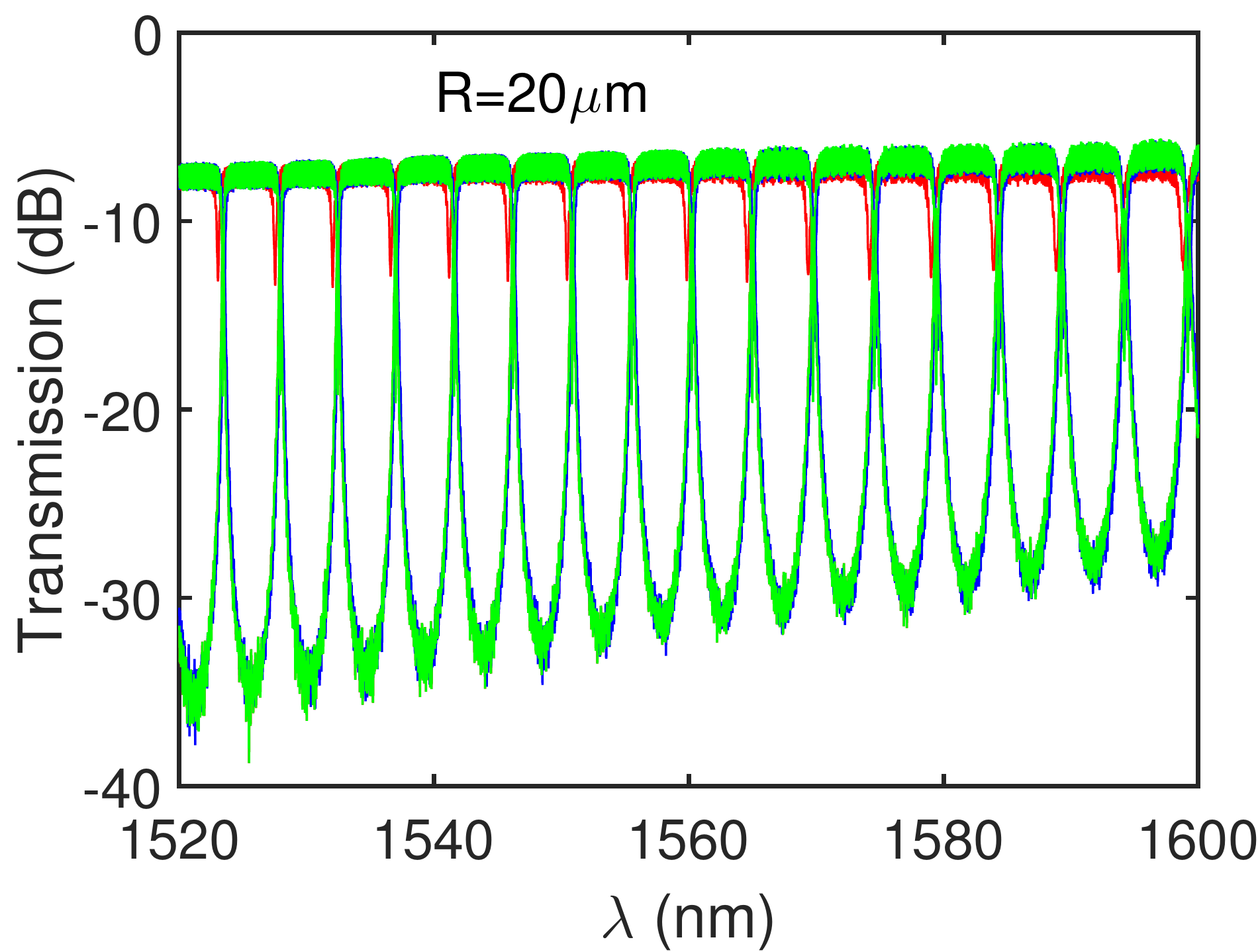}\\
  {\large (c)}\\
\includegraphics[width=0.4\columnwidth]{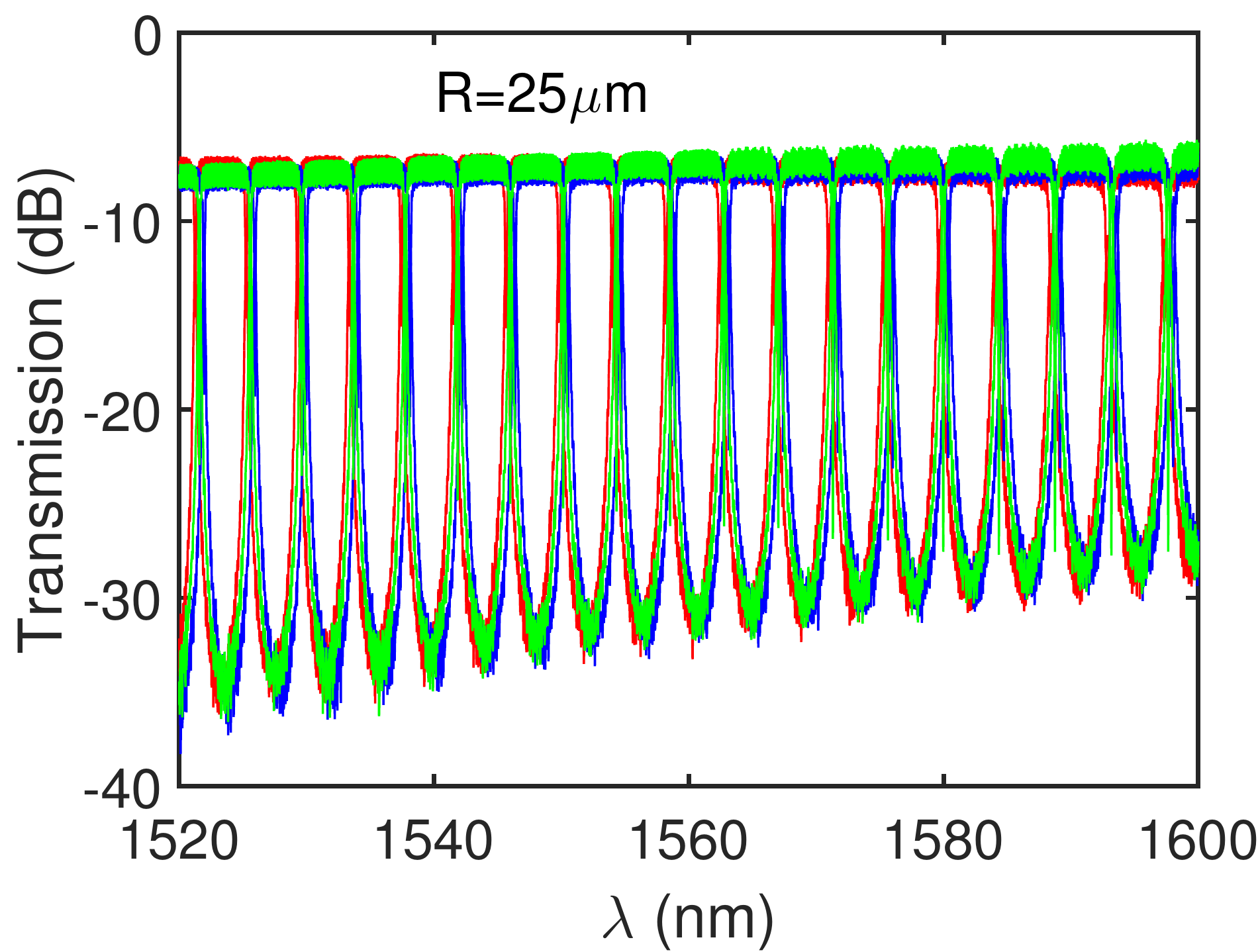}
\end{tabular}
\caption{Through and drop responses measured in the full C band. Plots (a), (b) and (c) correspond to radii of curvature of $R=15$ $\mu$m, $20$ $\mu$m and $25$ $\mu$m, respectively.  The results corresponding to the A-type geometry are shown in red, those of the B-type are plotted in blue, and green lines correspond to the C-type geometry. }\label{fig::resultados}
\end{figure}

\begin{figure}[!t]
\centering
\begin{tabular}{c}
  {\large (a)}\\
  \includegraphics[width=0.4\columnwidth]{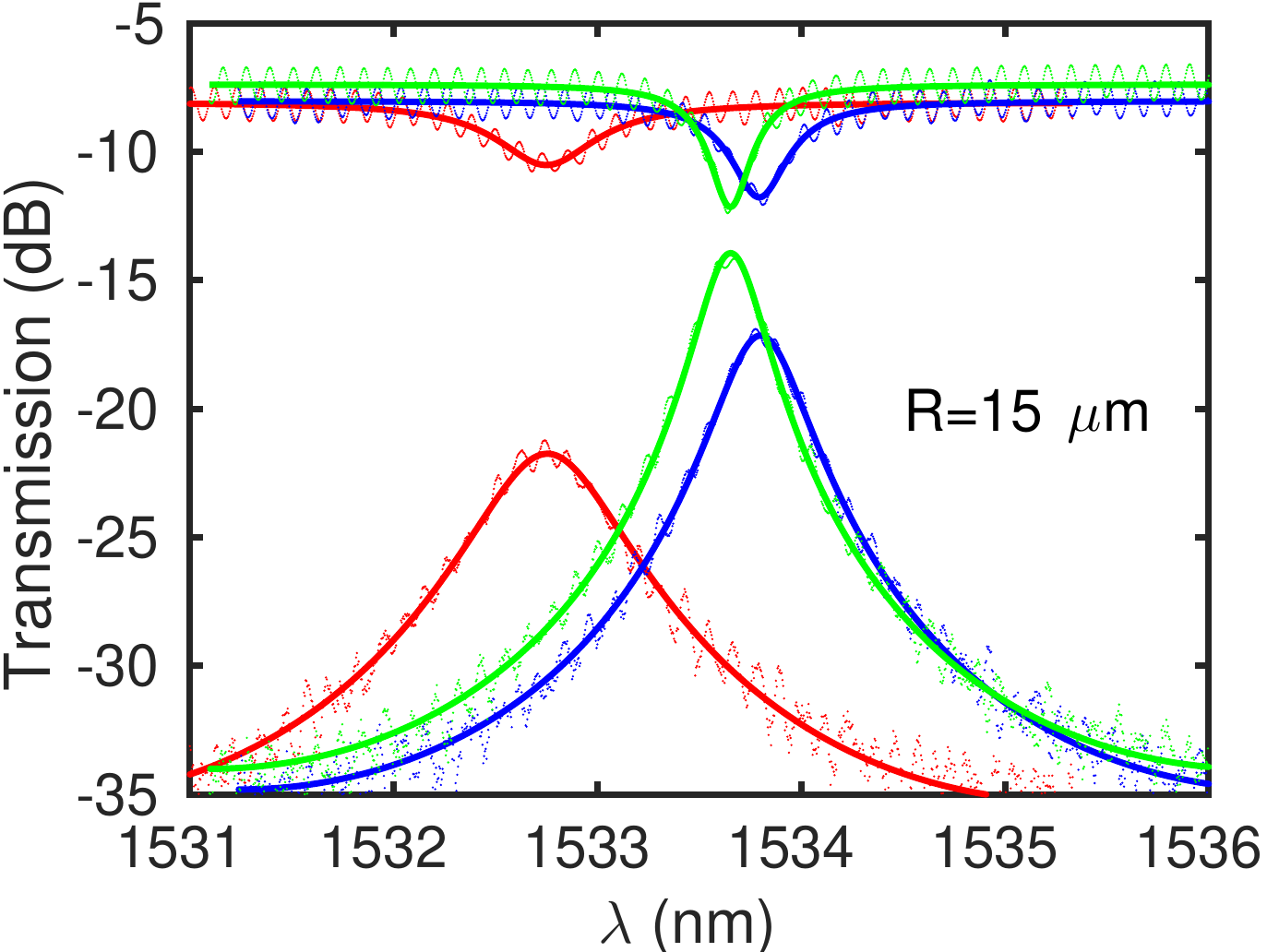}\\
  {\large (b)}\\
  \includegraphics[width=0.4\columnwidth]{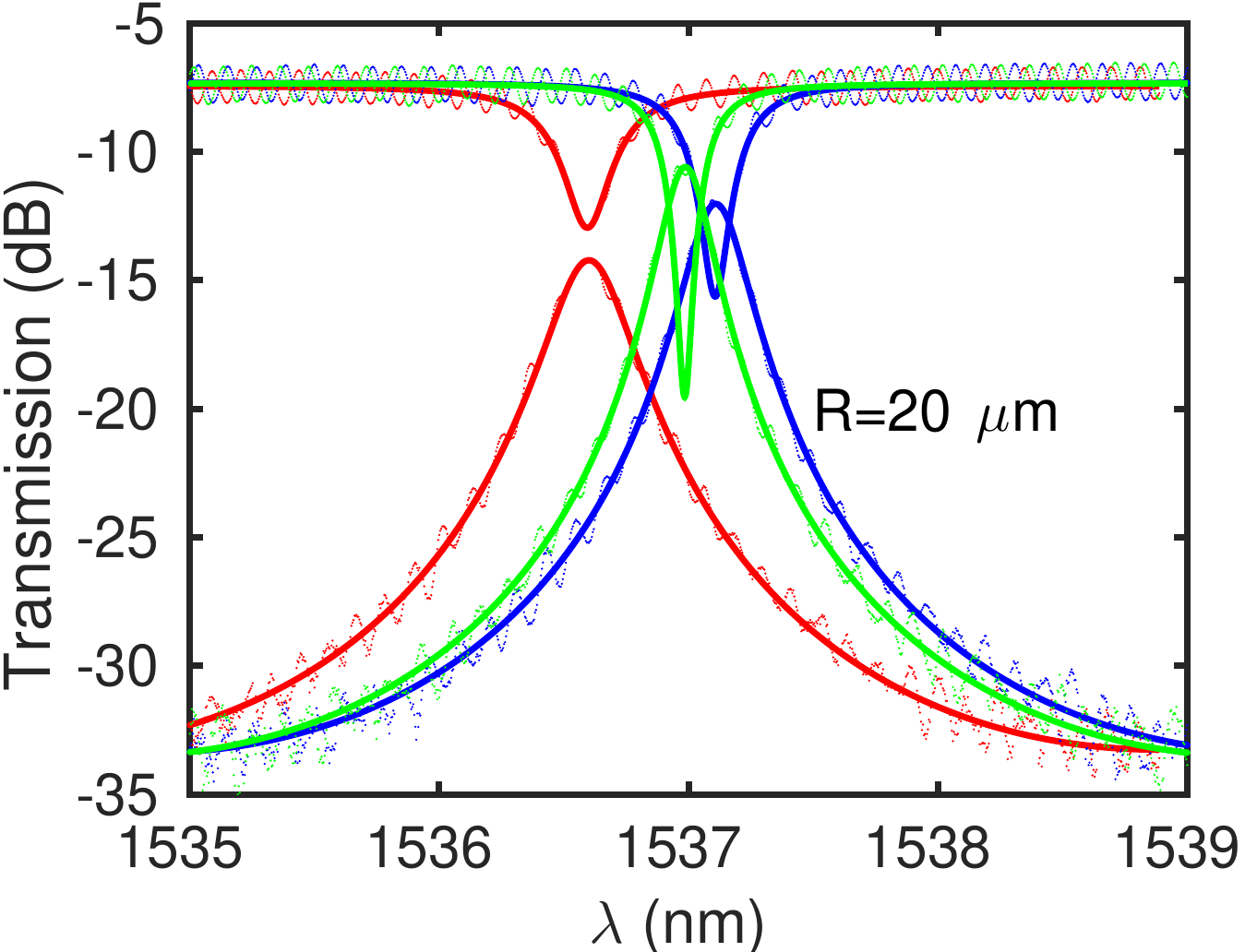}\\
  {\large (c)}\\
\includegraphics[width=0.4\columnwidth]{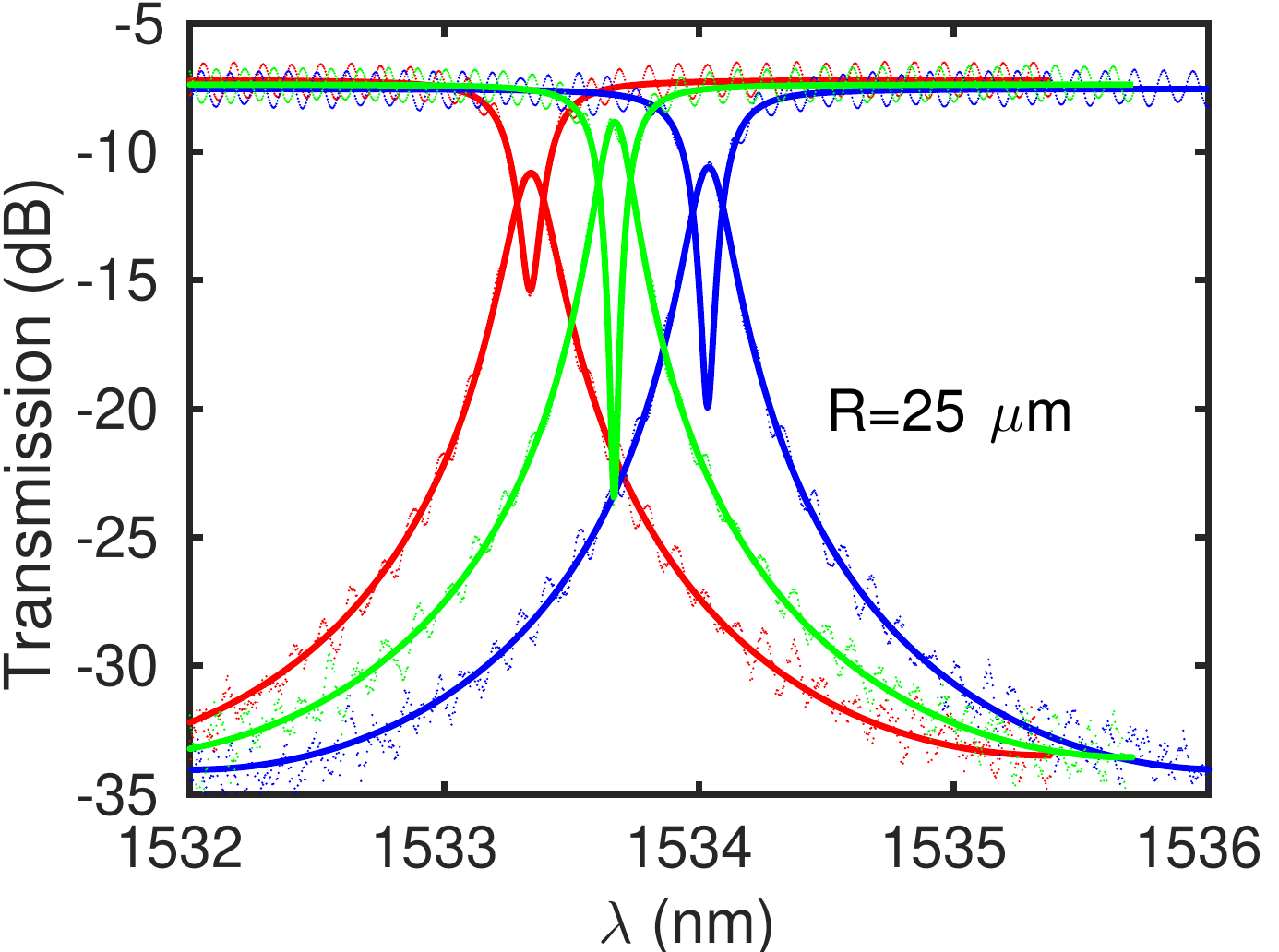}
\end{tabular}
\caption{Measured through and drop responses of the add-drop resonators at a resonance placed at the short wavelength end of the the C band. (a), (b) and (c) correspond to $R=15$ $\mu$m, $20$ $\mu$m and $25$ $\mu$m radii of curvature, respectively.  The results corresponding to the A-type geometry are shown in red, those of the B-type are plotted in blue, and green results correspond to the C-type geometry. Solid lines are obtained fitting the responses to spectral shapes of Eqs. \eqref{eq::respthrough} and \eqref{eq::respdrop}} \label{fig::resultadosS}
\end{figure}

\begin{figure}[!t]
\centering
\begin{tabular}{c}
  {\large (a)}\\
  \includegraphics[width=0.4\columnwidth]{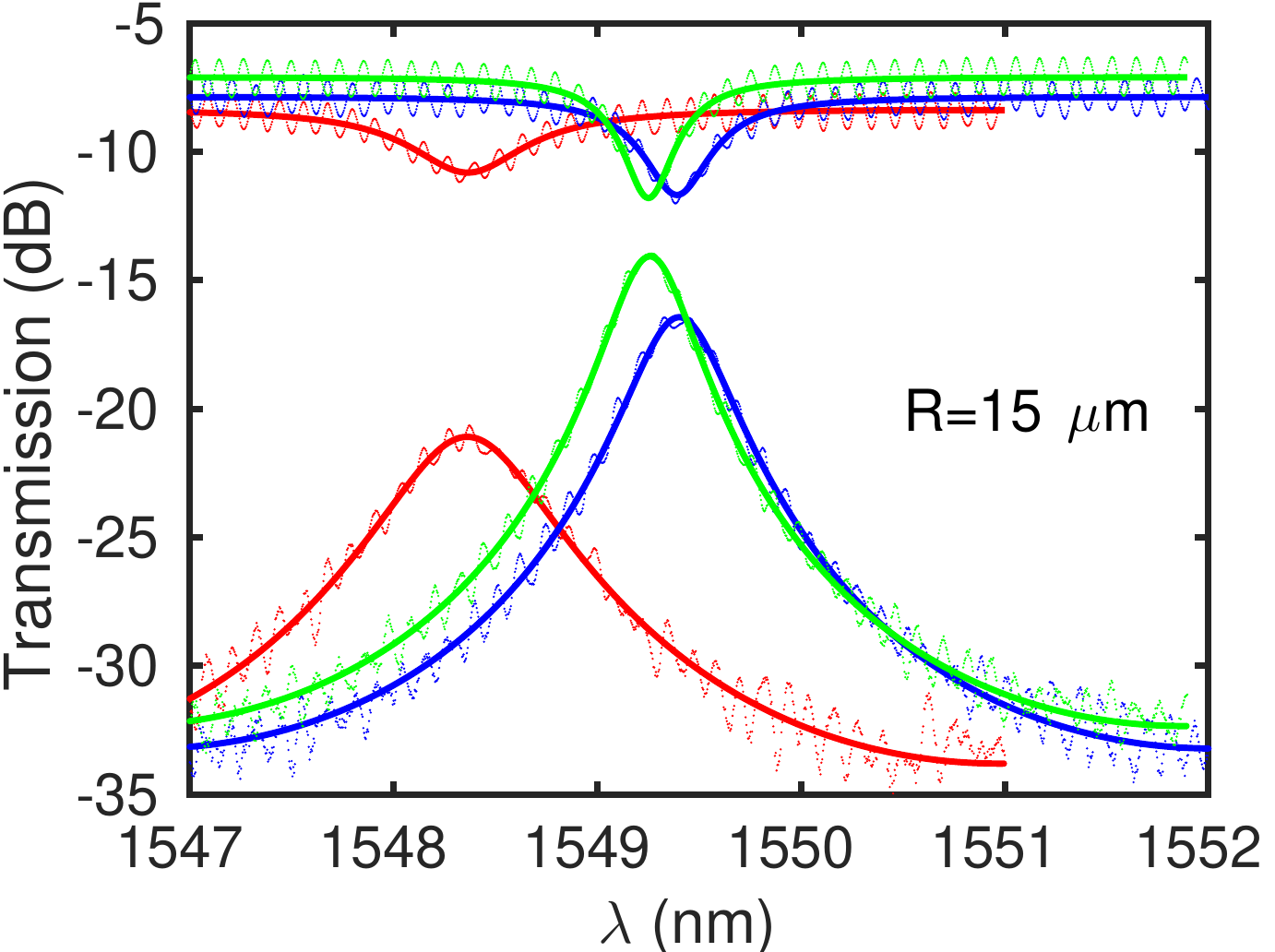}\\
  {\large (b)}\\
  \includegraphics[width=0.4\columnwidth]{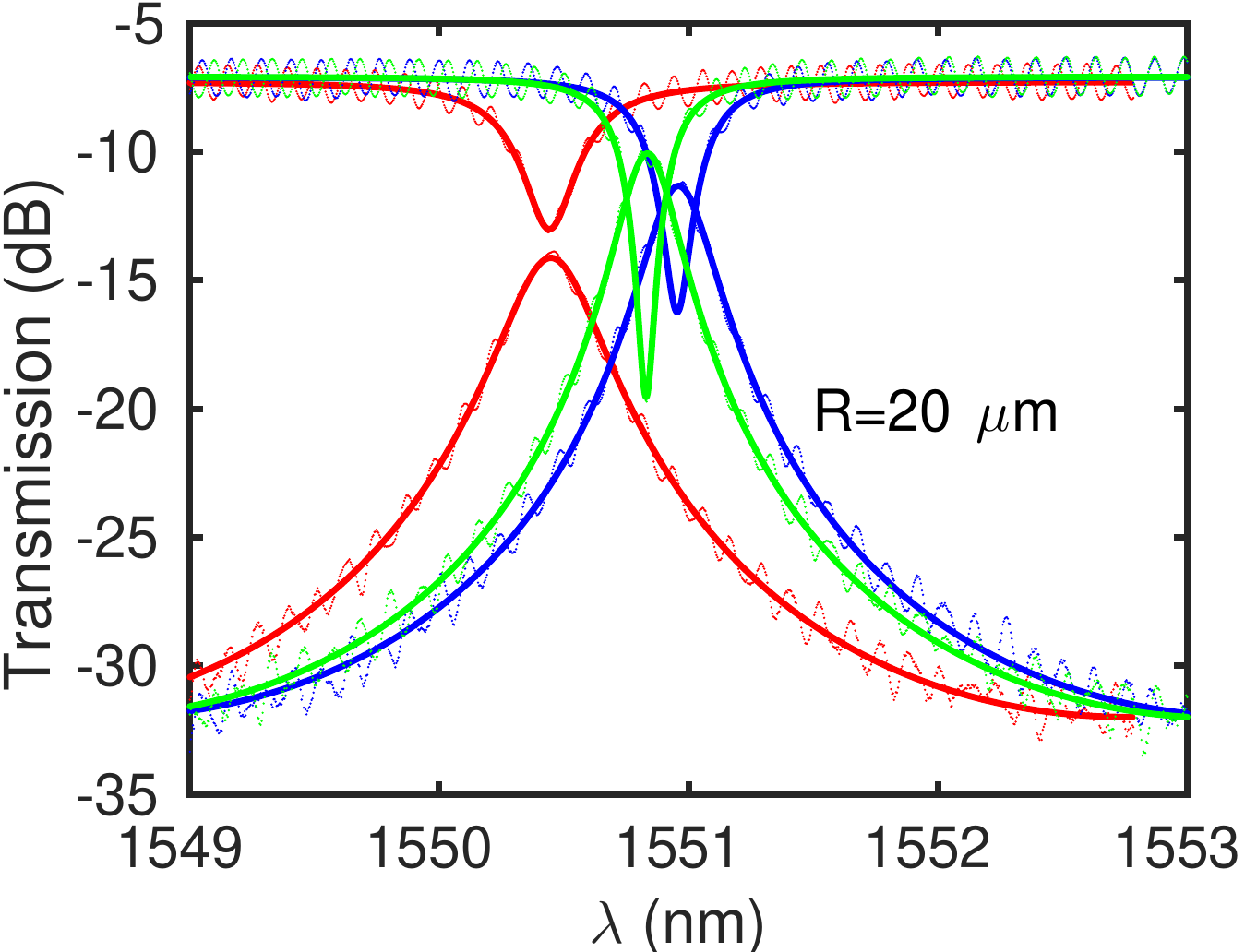}\\
  {\large (c)}\\
\includegraphics[width=0.4\columnwidth]{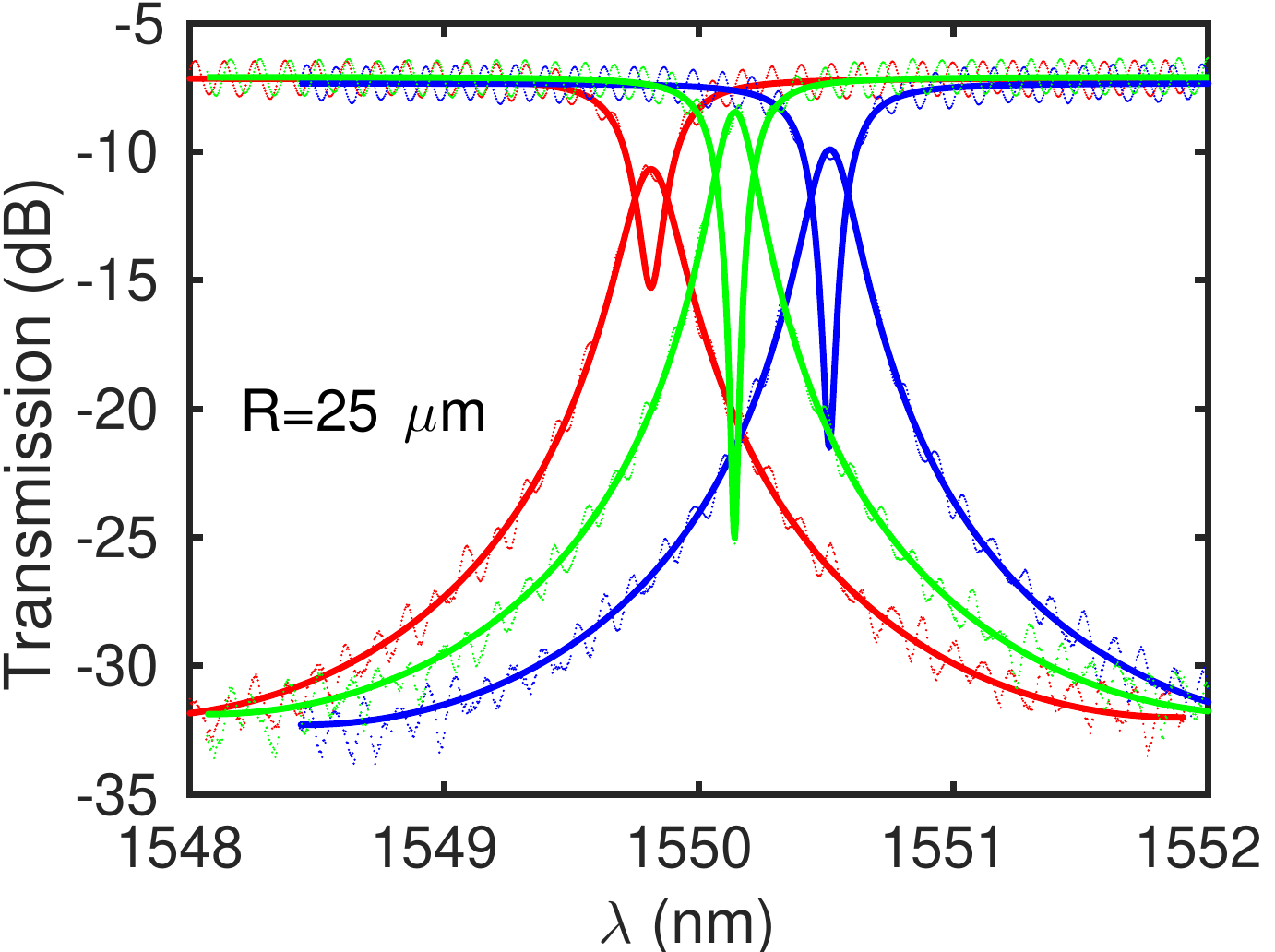}
\end{tabular}
\caption{Measured through and drop responses of the add-drop resonators at a resonance placed at the central region of the C band. (a), (b) and (c) correspond to $R=15$ $\mu$m, $20$ $\mu$m and $25$ $\mu$m radii of curvature, respectively.  The results corresponding to the A-type geometry are shown in red, those of the B-type are plotted in blue, and green results correspond to the C-type geometry. Solid lines are obtained fitting the responses to spectral shapes of Eqs. \eqref{eq::respthrough} and \eqref{eq::respdrop}} \label{fig::resultadosC}
\end{figure}

\begin{figure}[!t]
\centering
\begin{tabular}{c}
  {\large (a)}\\
  \includegraphics[width=0.4\columnwidth]{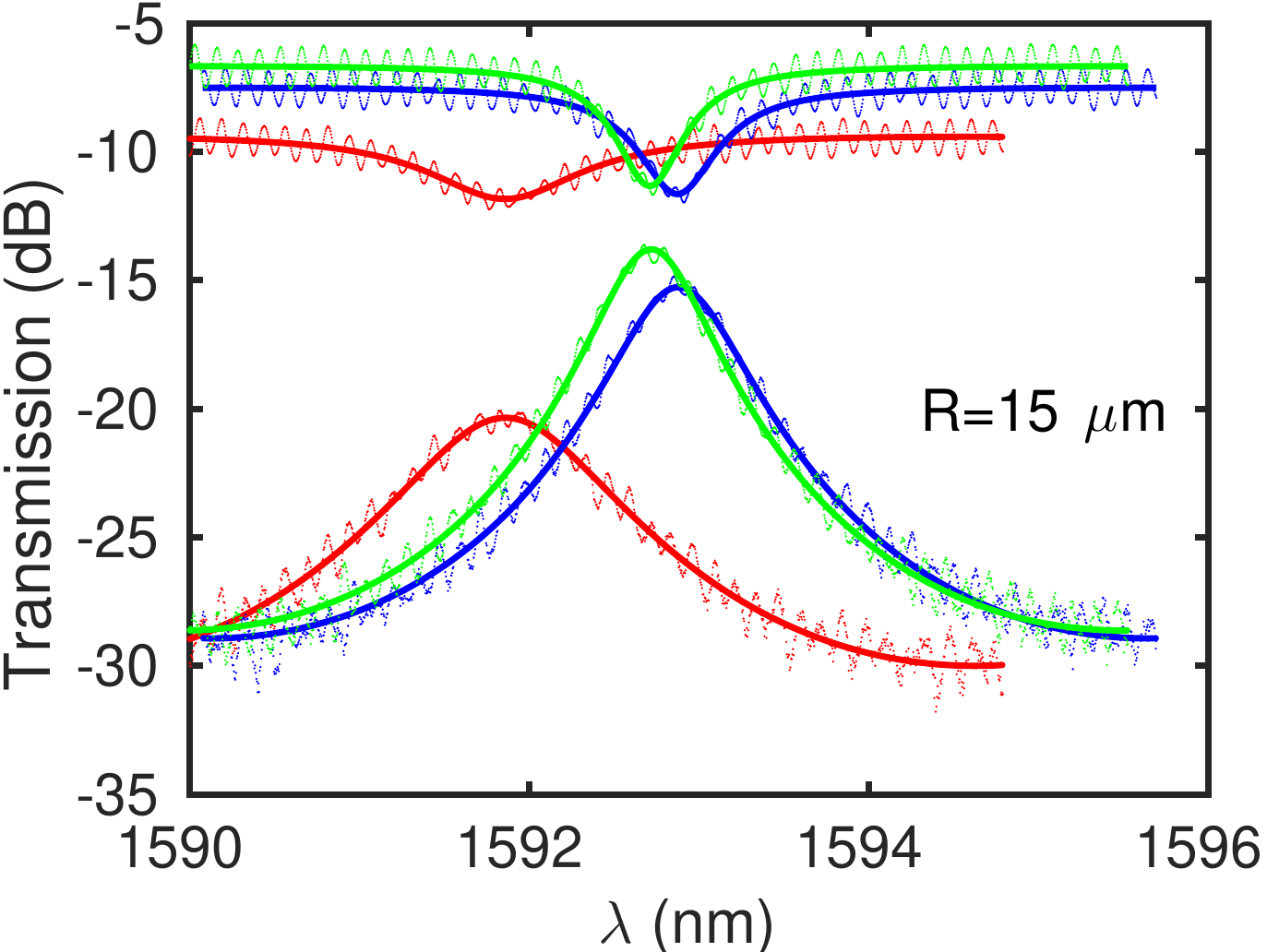}\\
  {\large (b)}\\
  \includegraphics[width=0.4\columnwidth]{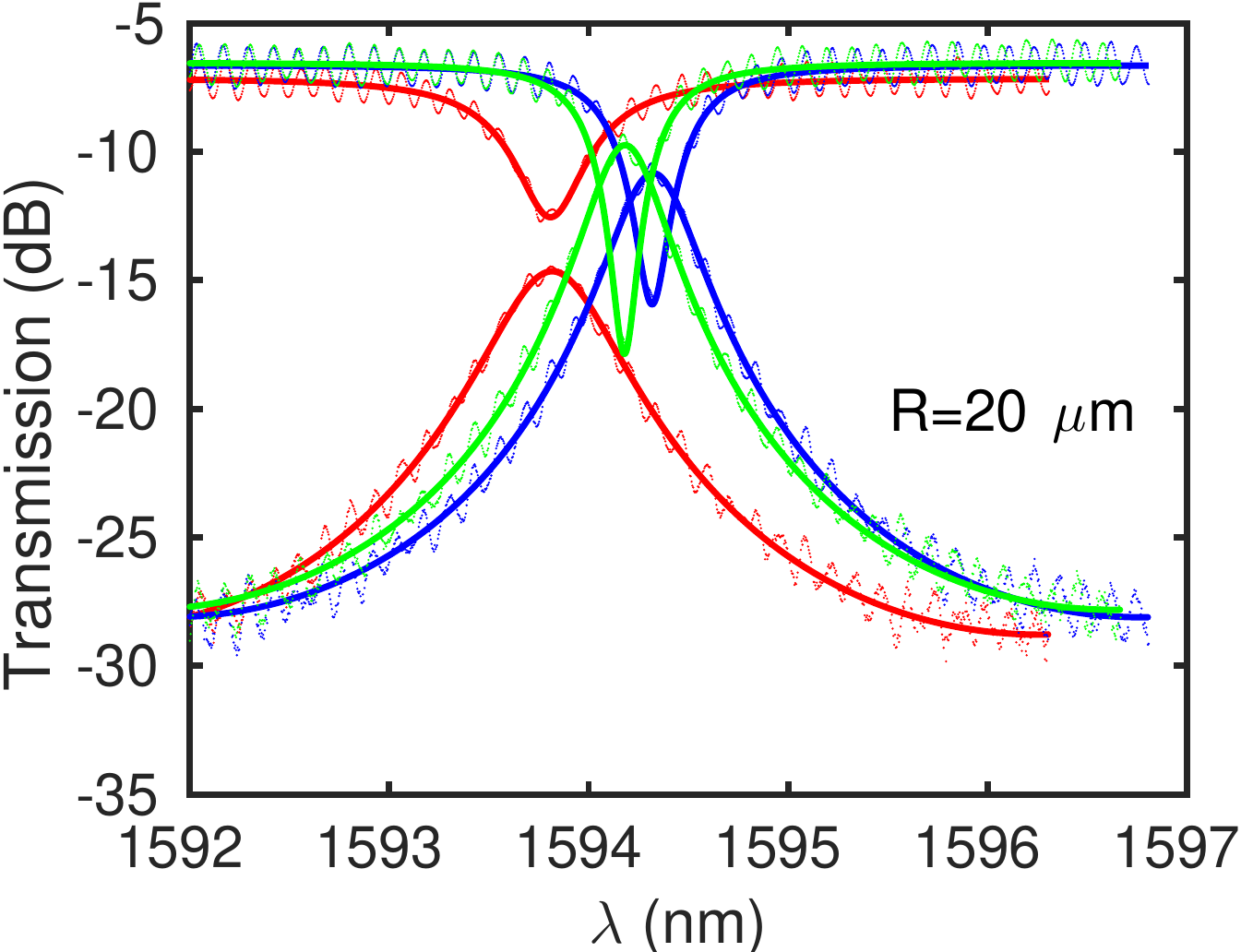}\\
  {\large (c)}\\
\includegraphics[width=0.4\columnwidth]{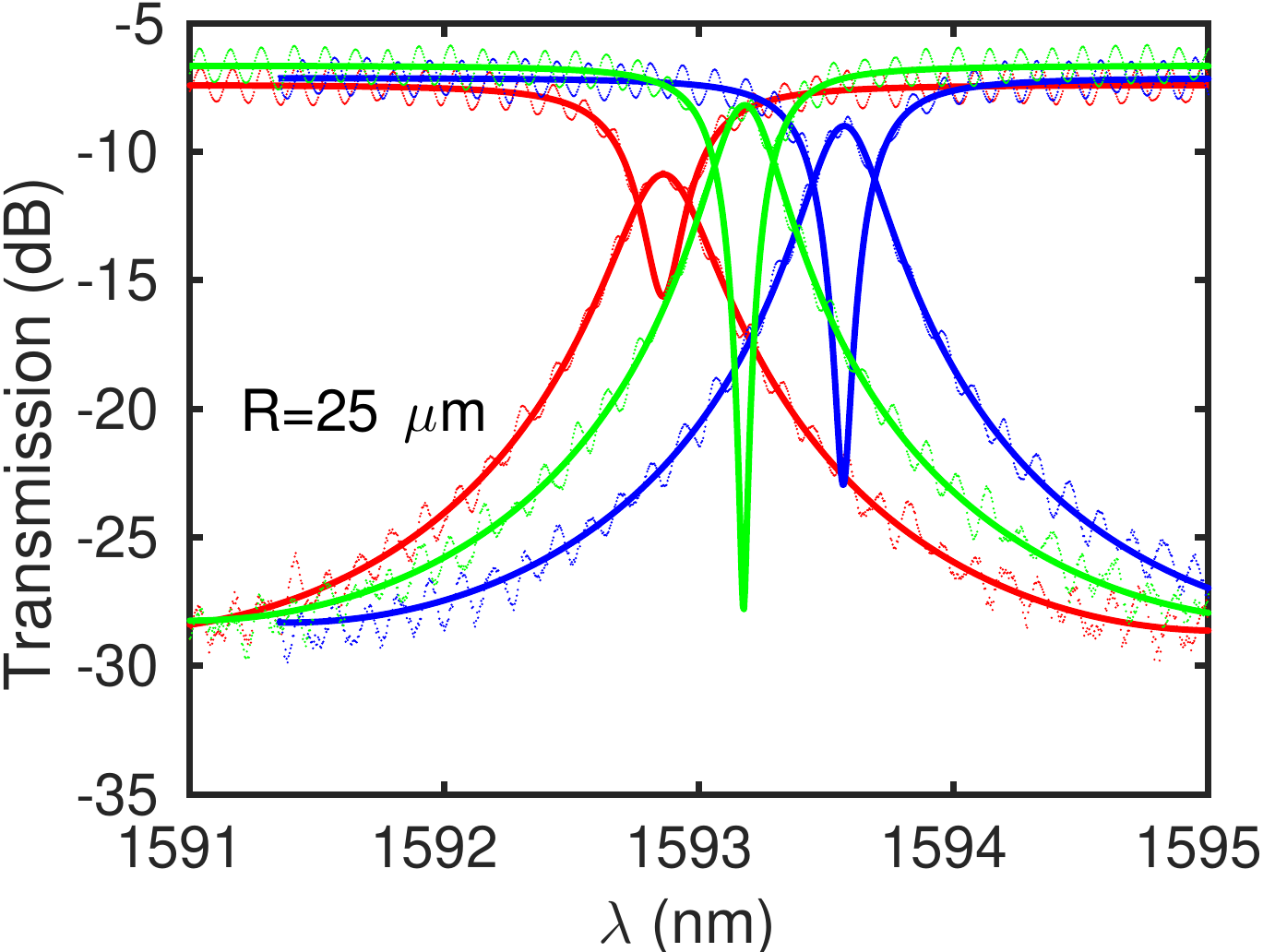}
\end{tabular}
\caption{Measured through and drop responses of the add-drop resonators at a resonance placed at the short wavelength end of the the C band. (a), (b) and (c) correspond to $R=15$ $\mu$m, $20$ $\mu$m and $25$ $\mu$m radii of curvature, respectively.  The results corresponding to the A-type geometry are shown in red, those of the B-type are plotted in blue, and green results correspond to the C-type geometry. Solid lines are obtained fitting the responses to spectral shapes of Eqs. \eqref{eq::respthrough} and \eqref{eq::respdrop}} \label{fig::resultadosL}

\end{figure}

\begin{sidewaystable}
\centering
\begin{tabular}{l||c||c||c}
  \hline
  \hline
  &$R=15$ $\mu m$ & $R=20$ $\mu m$ & $R=25$ $\mu$m\\
  \hline
  \hline
  &\begin{tabular}{ccc}
    $EF_{BA}$&$EF_{CB}$&$EF_{CA}$
  \end{tabular}&
   \begin{tabular}{ccc}
    $EF_{BA}$&$EF_{CB}$&$EF_{CA}$
   \end{tabular}&
   \begin{tabular}{ccc}
    $EF_{BA}$&$EF_{CB}$&$EF_{CA}$
  \end{tabular}\\
   \hline
   $1531-1539$ nm&
\begin{tabular}{ccc}
    $1.87$&$1.42$&$2.65$
  \end{tabular}&
   \begin{tabular}{ccc}
    $1.81$&$1.99$&$3.60$
   \end{tabular}&
   \begin{tabular}{ccc}
    $1.75$&$1.81$&$3.16$
   \end{tabular}\\
   \hline
   $1547-1553$ nm&
\begin{tabular}{ccc}
    $1.84$&$1.34$&$2.47$
  \end{tabular}&
   \begin{tabular}{ccc}
    $2.02$&$1.70$&$3.42$
   \end{tabular}&
   \begin{tabular}{ccc}
    $2.33$&$1.72$&$4.00$
   \end{tabular}\\
  \hline
  $1590-1597$ nm&
\begin{tabular}{ccc}
    $2.10$&$1.19$&$2.51$
  \end{tabular}&
   \begin{tabular}{ccc}
    $2.21$&$1.37$&$3.03$
   \end{tabular}&
   \begin{tabular}{ccc}
    $2.81$&$2.07$&$5.82$
   \end{tabular}
    
\end{tabular}
\caption{Q enhancement factors for the three values of $R$ considered in this work and and resonance frequencies distributed along the C band.  $EF_{JI}$ denotes the enhanced factor when the design is modified changing geometry I to geometry J}\label{table}
\end{sidewaystable}

Radiation loss can be modified by adding an external waveguide section, as shown in Fig. \ref{fig::estructuras} B and C.  The radiation quenching properties of coupled asymmetric curved waveguides have been described in detail in \cite{radiacion}, using highly accurate 3D vector modal calculations, in terms of the variation of the properties of modes propagating in the structure as a function of the waveguide separation $s$ and the width of the exterior waveguide $w_e$ when the width of the main waveguide $w$ is kept fixed and the radius of curvature $R$ is varied.  In the following paragraphs, we provide a precise explanation of this effect.

We consider first a straight coupler with $R\to\infty$.  In the symmetrical case, $w_e=w$, we have a modal degeneracy at $s\to \infty$ with each mode occupying one of the cores of the well-separated and, therefore, uncoupled waveguides.  As the two waveguides become closer, the degeneracy is broken due to mutual coupling with the result of a splitting of the propagation constants of the two super-modes.  Two branches, with a separation that grows as $s$ decreases, can be observed when the modal indices are plotted as a function of $s$.  These two branches correspond, respectively, to odd and even modal fields that have intensities evenly split between the two waveguide cores at small $s$ in the symmetrical case.  When the modal indices are evaluated over the $(w_e,s)$ plane, this splitting results in the creation of two sheets; the lower sheet corresponds to the odd mode and the upper sheet to the even mode.  For $w_e<w$, the even mode is predominantly localized at the main waveguide, while it is the odd mode the one that is localized at the main waveguide for $w>w_e$ \cite{radiacion}.

As $w_e$ is increased even further, new super-modes are supported by the two-waveguide coupler.  These modes show an increasingly number of maxima and/or minima of the transverse field within the rightmost waveguide of width $w_e$. Figure \ref{fig::EIM} (a) shows the effective indices of the propagating modes as $w_e$ and $s$ are varied, calculated using the effective index method \cite{radiacion}.  At most locations in the $(w_e,s)$ domain one of the sheets is nearly coincident with the $n=n_0$ plane, where $n_0$ is the effective index of the single waveguide of width $w=1$ $\mu m$.  Precisely, this is true except at the smallest values of $s$ and the transition regions where there is a modification of which modal sheet is the closest to the $n_{eff}=n_0$ plane. This sheet corresponds the mode predominantly guided in the main (leftmost) waveguide at a given value of $s$ and $w_e$.  In the following discussion, we will refer to this mode as the principal mode.  Figure \ref{fig::EIM} (b) displays the ratio of the  power carried in the core of the rightmost waveguide $P_2$ to that carried in the core of the main waveguide $P_1$ by the principal mode at each value of $s$ and $w_e$.  The results of Figure \ref{fig::EIM} (b) show that  this mode is highly confined in the core of the first waveguide and $P_2/P_1\to 0$ for most values of $s$ and $w_e$.  On the other hand, there are clearly visible regions where this condition is not fulfilled and $P_2/P_1$ takes large values. These regions of poorer confinement of the principal mode in the main waveguide coincide with those of small values of $s$ or the transitions of the modal sheet corresponding to the principal mode that are associated to larger deviations of $n_{eff}$ from $n_0$ in Figure \ref{fig::EIM} (a), as described above.

Curving the asymmetric coupler induces a distortion of the effective transverse refractive index profile seen by the propagating field \cite{hiremath}.  In turn, this produces a deformation of the modal sheets described above, the effective degeneracy condition, and the transition regions associated to the appearance of new quasi-modes of the bent asymmetric coupler \cite{radiacion}.  Moreover, the associated regions with high and low confinement of the principal mode in the main waveguide become accordingly distorted.  For the small values of $R$ required in a photonic integrated circuit (PIC) design, the first odd mode becomes the principal mode over a wide domain in the $(s,w_e)$ plane extending even in the $w_e<w$ region\cite{radiacion}.  The imaginary parts of the complex propagation constants of the quasi-modes of a curved waveguide are defined by their coupling to the radiation field associated to the waveguide curvature \cite{hiremath}. This coupling will be highly quenched if the modal field is well-confined in the innermost main waveguide and, on the contrary, an enhancement of radiation loss will occur when a larger fraction  of the total power is carried by the outermost waveguide, therefore facilitating the linkage to the radiation field. 

By studying the radiation properties of the principal mode of a curved asymmetric coupler as the parameters $w_e$ and $s$ are varied, one finds that the radiation loss described by the imaginary part of the modal effective index varies in a regular manner \cite{radiacion} that is dependent on the radius of curvature, as shown in Figure \ref{fig::contornos} where the results have been obtained using a highly accurate full vector numerical method.  As explained above, the contours shown in this figure are the result of the deformation of the domains shown in Figure \ref{fig::EIM} (b) as the coupler bends. The results of Figure \ref{fig::contornos} permit to select points in the $(s,w_e)$ plane that give designs with reduced radiation loss.  This type of variations of the radiation losses had also been found in the 2D case using finite-differences time-domain (FDTD) calculations \cite{chamorro}. As expected, the change in the imaginary part of the effective modal index is accompanied by a change in the modal field distribution such that transverse intensity distribution is more localized at the waveguide when $n_i$ decreases and expands outwards in the transverse plane as $n_i$ increases.    

Curving a single waveguide with a finite radius $R$, besides of causing a change in the effective index and making it complex, produces a shift of the modal field intensity distribution mainly along the radial direction \cite{radiacion}.   The modification of the modal field and its impedance at the curved sections creates a discontinuity at the straight-bent interfaces that generates localized radiation beamed along the direction of the incoming optical field \cite{chamorro}.  This radiation can be reduced by shifting the relative central positions of the straight and bent waveguides at their junction so as to maximize the mode overlap between the mode field distribution in the two sections.  This is the so-called lateral shift technique \cite{kitoh} and its use is illustrated in Fig. \ref{fig::estructuras}.  The implementation of this technique in racetrack GaAs-AlGaAs micro-resonators was demonstrated in \cite{van}.

The transmission of the input optical intensity to the through and drop ports ($T_t$ and $T_d$, respectively) in Fig. \ref{fig::estructuras} is given by \cite{bogaerts}
\begin{eqnarray}
T_t =r^2\dfrac{a^2-2 a \cos \phi +1}{1-2 r^2 a \cos\phi +(r^2a)^2}\label{eq::respthrough}\\
T_d =\dfrac{(1-r^2)^2 a}{1-2 r^2 a \cos\phi +(r^2a)^2}, \label{eq::respdrop}
\end{eqnarray}
where $\phi=n_g 2 \pi L/\lambda$ and $a$ is the total attenuation factor of the optical field in one round-trip along the cavity including the bending losses, the coupling losses between straight and bent sections, and the intrinsic propagation losses.  This latter contribution can be neglected in the fabricated structures, since the total attenuation in the straight waveguide sections is below $0.5$ dB/cm.   The group index at wavelength $\lambda$ is given by
\begin{equation}
  n_g=n_{eff} - \lambda \dfrac{d n_{eff}}{d\lambda},
\end{equation}
with $n_{eff}$ the effective modal index.

The total Q factor is given  \cite{bogaerts} by the expression
\begin{equation}
  Q=\dfrac{\pi n_g L r \sqrt{a}}{\lambda_{res}\left(1-r^2a\right)},\label{eq::q}
\end{equation}
whereas the intrinsic Q-factor, corresponds to the case where the resonator is decoupled form the input and output access waveguides $r\to 1$ and is directly related with the round-trip loss $a$ as 
\begin{equation}
  Q_i=\dfrac{\pi n_g L \sqrt{a}}{\lambda_{res}\left(1-a\right)},\label{eq::qi}
\end{equation}
where $\lambda_{res}$ is the resonance wavelength.

\section{Design}\label{sec::design}

Our design strategy aims to the minimization of the total round-trip loss by optimizing both the continuous radiation quenching asymmetric coupler and the lateral offsets for the localized radiation at discontinuities.

In the calculations, the refractive index of silicon dioxide \cite{malitson} has been taken as $n_{SiO_2}=1.4440$ at $\lambda=1550$ nm and that of silicon nitride \cite{luke} $n_{Si_3N_4}=1.9963$.  The width of the main waveguide is taken as $w=1$ $\mu$m and the height of all the silicon nitride sections $h=300$ nm.

The principal mode of the curved asymmetric coupler formed by the addition of an exterior, radiation quenching, curved section to the main waveguide in Fig. \ref{fig::estructuras} B has been traced over a wide domain in the $(w_e,s)$ plane. The corresponding modal losses, as obtained with full
3D vector calculations using the \emph{wgsm3d} software package \cite{krause,wgms3d} expressed as $log_{10}\left(n_i\right)$, where $n_i$ is the imaginary part of the complex modal effective index are shown in Fig. \ref{fig::contornos} for $R=15$ $\mu$m, $R=20$ $\mu$m and $R=25$ $\mu$m.

Fig. \ref{fig::contornos} displays, for the three values of the bend radius considered, a main broad parameter region adequate for radiation quenching at small values of both the waveguide separation $s$ and width of the exterior coupled waveguide $w_e$. In this region, the principal mode coincides with the first odd-order mode of the curved asymmetric coupler.  The fact that this favorable design zone of the two-dimensional parameter space is broad, with smooth variations of the optimization target,  contributes to the broadband character of the designs and their resilience to fabrication tolerances.  As $R$ increases, the main valley of the curves in Fig. \ref{fig::contornos} shifts to larger values of $s$ and $w_e$, as expected.  In our design we have used values of $s=1.3$ $\mu$m, $s=1.7$ $\mu$m and $s=1.9$ $\mu$m for $R=15$ $\mu$m, $R=20$ $\mu$m and $R=25$ $\mu$m, respectively. The values of the exterior waveguide width have been set to $w_e=600$ nm, $w_e=600$ nm and $w_e=700$ nm for $R=15$ $\mu$m, $R=20$ $\mu$m and $R=25$ $\mu$m, respectively.

The design of the lateral shifts at the straight-bent discontinuities has been based on the maximization of the mode overlap of the guided modes in the two sections \cite{chamorro,polarizadores,radiacion}. The calculated optimal offset values are $l_{off}=110$ nm, $90$ nm, and $70$ nm for $R=15$ $\mu$m, $20$ $\mu$m and $25$ $\mu$m, respectively.

All the designs have targeted an operation wavelength of $1550$ nm. The self-coupling coefficients of the evanescent couplers are affected by the introduction of lateral offsets at the straight-bent discontinuities when straight input and output waveguides are used \cite{chamorro}.  The modification of the value of $r$ would affect the total Q factor and would hinder a visual appreciation from the the device responses of the effect of the proposed geometry modifications on the intrinsic Q factor. Therefore, the coupling sections in our devices are kept far from the terminations of the straight arms of the racetrack in order to preserve the values of $r$ in cases $A$, $B$, and $C$.

\section{Results}

The silicon nitride chip was fabricated at the Instituto de Microelectrónica de Barcelona, Centro Nacional de Microelectrónica (IMB-CNM), CSIC \cite{VLC} through a Multi Project Wafer approach offered by VLC Photonics.  The designed devices have been fabricated in a $5$mm$\times 5$mm silicon nitride integrated circuit.  Figure \ref{fig::imagenes} displays micrographs of two of the fabricated devices with $R=15$ $\mu$m corresponding to geometries $A$ and $B$.  

The manufactured resonators have been characterized in the C band.   The through and drop responses of the racetrack micro-resonators have been measured in the range between $1520$ nm and $1600$ nm with a spectral resolution of $20$ pm  using a AQ6370D Yokogawa Optical Spectral Analyzer.  The input and output couplings have been performed using objectives with a mode field diameter of approximately $2.5$ microns and TE polarization filters.  The coupling losses are estimated to be between $2.75$ and $3$ dB per facet.

Figures \ref{fig::resultados} (a), (b) and (c), show the through and drop responses for the fabricated rings with radii $15$, $20$ and $25$ $\mu$m, respectively. Red, blue and green traces correspond, respectively, to the configurations $A$, $B$ and $C$ of Figure \ref{fig::estructuras}.  Figures \ref{fig::resultadosS}, \ref{fig::resultadosC} and \ref{fig::resultadosL} display the measured through and drop responses of the devices (dotted lines) at resonances placed at three different locations within the C band: close the short wavelength end, in its central region, and near the long wavelength end.  The coupling losses from the measurement apparatus to the PIC are also included in these plots. This is the reason why the baselines for the drop and through responses of different ring resonators are very similar in Figures \ref{fig::resultados}, \ref{fig::resultadosS}, \ref{fig::resultadosC} and \ref{fig::resultadosL}. The measured responses exhibit a rippling that is due to the etalon effect due to reflections at the end facets of the integrated circuits. Solid lines display the fitting of the measured data to the through \eqref{eq::respthrough} and drop \eqref{eq::respdrop} responses obtained using Matlab's linear programming solver \emph{fminsearch}.

 In all the cases, the improvement of the Q-factors from all the cases corresponding to the conventional (A, red) geometry to that including the radiation quenching sections (B, blue) and the lateral offsets (C, green) is evident from the plots.  Since the design preserves the value of $r$, as discussed in Section \ref{sec::design}, this improvement can be directly assigned to the reduction of the total round-trip loss $a$ and the associated enhancement of the internal Q of the resonators.

The total Q factor given in Eq. \eqref{eq::q} is a design parameter that can be controlled by the choice of the value of $r$, for instance, in a filter arrangement.  In any case, the performance of the optical system will be limited by the propagation losses that determine the insertion loss of the filter.  Therefore, when using the Q factor as a figure of merit, it is more adequate to use the intrinsic Q factor, Q\textsubscript{i}, of Eq. \eqref{eq::qi}, which is the maximum attainable Q factor for a given resonator length at a given wavelength and it is limited by the propagation loss associated to the curved waveguide sections.
 
 The results of the linear programming fits permit to determine de values of $a$ and the calculation of the intrinsic Q enhancement factors, defined as the ratio of the intrinsic Q factor of the two compared geometries
 \begin{equation}
   EF_{JI}=\dfrac{Q_J}{Q_I}.
   \end{equation}
The results are shown in Table \ref{table}.  The magnitudes of the enhancement factors obtained due to the introduction of radiation quenching sections are comparable, but larger in almost all the cases cases, to those due to the presence of lateral offsets.  

  The largest values of Q\textsubscript{i} obtained in the measurements of Figures \ref{fig::resultadosS}, \ref{fig::resultadosC} and \ref{fig::resultadosL}: those that correspond to designs that include both Q-enhancement measures, are $7.72\times 10^3$, $3.08\times 10^4$ and $6.34\times 10^4$ for $R=15$ $\mu$m, $R=20$ $\mu$m, and $R=25$ $\mu$m,  respectively, in the shorter wavelength end of the C band; $6.29\times 10^3$, $2.68\times 10^4$ and $6.69\times 10^4$ for $R=15$ $\mu$m, $R=20$ $\mu$m, and $R=25$ $\mu$m,  respectively, at the center of the C band,  and $4.14\times 10^3$, $2.44\times 10^4$ and $6.55\times 10^4$, at the long wavelength edge.   It is important to stress, though, that Q\textsubscript{i} scales linearly with the total resonator length, as given by Eq. \eqref{eq::qi}, since the low values of intrinsic propagation loss permits to safely neglect the losses in the straight sections of the proposed geometries, and much larger values of  Q\textsubscript{i} can be realized by a simple elongation of the straight waveguide sections in the racetrack geometry. The values of the total propagation loss factor $a$, attributable to radiation losses in the curved waveguides and the couplings of straight-bent sections corresponding to the previous values of Q\textsubscript{i}, are $0.887$, $0.966$, and $0.982$, at the shorter wavelength band edge, $0.864$, $0.962$, and $0.983$, in the band center, and $0.805$, $0.932$ and $0.983$ at the long wavelength end of the C band.  These values of $a$ will ultimately limit the performance of any racetrack resonator within the design restrictions of this work in their respective wavelength region for each value of $R$.

\section{Conclusion}

Q-enhanced racetrack microresonators have been fabricated in the silicon nitride platform and their responses have been characterized.  The measurements performed confirm a significant reduction of the radiation losses in these structures, as predicted by previous theoretical studies.  The micro-resonator geometries experimentally demonstrated permit reduced sized implementations when designed for the same operation conditions as their conventional counterparts and, therefore, they contribute to an increment of the integration scale in this platform.  Furthermore, in the silicon nitride platform, restriction on the radii of curvature
also apply to all the curved optical connections in the layout of any PIC in order to limit the effects of radiation loss.  The Q-enhanced design strategy devised for micro-resonators can be used in any curved waveguide section used to carry the optical signals in the circuit and it would allow for a reduction of the permitted radii of curvature, also contributing to the increase of the integration scale in this platform.  Besides the radiation quenching properties of the geometry modifications employed in the Q-enhanced geometries, they also hold strong polarization dependent properties that can be exploited in the design of polarization control devices \cite{polarizadores}.

\section*{Acknowledgements}

This work has been supported by Junta de Castilla y Le\'on, Project No. VA296P18.

Micrographs where performed at the Unidad de Microscop\'{\i}a of Parque Científico  at UVa.


\begin{thebibliography}{99}

\bibitem{bogaerts} W. Bogaerts, P. DeHeyn, T. Van Vaerenbergh, K. De Vos, S. K. Selvaraja, T. Claes, P. Dumon, P. Bienstaman, D. Van Thourhout, and R. Baets, ``Silicon microring resonators,'' Laser Photonics Rev., vol. 6, no. 1, pp. 47--73, 2012.


\bibitem{little1} B.E. Little, S.T. Chu, H.A. Haus, J. Foresi, and J.-P. Laine, ``Microring Resonator Channel Dropping Filters,'' J. Lightwave Technol., vol. 15, pp. 998-1005, June 1997

\bibitem{little2} B.E. Little, S.T. Chu, P.P. Absil, J.V. Hryniewicz, F.G. Johnson, F. Sieferth, D. Gill, V.Van, O. King and M. Trakalo, ``Very High-Order Microring Resonator Filters for WDM Applications,'' IEEE Photon. Technol. Lett., vol. 16, pp. 2263--2265, Oct. 2004.

\bibitem{chamorro11} P. Chamorro-Posada, F. J. Fraile-Pelaez, and F.J. Diaz-Otero, ``Micro-Ring Chains With High-Order Resonances,'' J. Lightwave Technol., vol. 29, pp. 1514--1521, May 2011

\bibitem{almeida} V.R. Almeida and M. Lipson, ``Optical bistability on a silicon chip,'' Opt. Lett., vol. 29, pp.  2387--2389, Oct. 2004.

\bibitem{vos} K. De Vos, J. Girones, S. Popelka, E. Schacht, R. Baets, and P. Bienstman, ``SOI optical microring resonator with poly(ethylene glycol) polymer brush for label-free biosensors applications,'' Biosens. Bioelectron., vol. 24, pp. 2528--2533, 2009.

\bibitem{xu} Q. Xu, B. Schmiddt, s. Pradhan, and M. Lipson, ``Micrometre-scale silicon electro-optic modulator,'' Nature, vol. 435, pp. 325--327, May 2005.

\bibitem{scheuer} J. Scheuer, G.T. Paloczi, J.K.S. Poon, and A. Yariv, ``Coupled resonator optical waveguides: Towards Slowing and Storing of Light,'' Opt. Photon. News, vol. 16, pp. 36--40, Feb. 2005.

\bibitem{chamorro09} P. Chamorro-Posada, and F.J. Fraile-Pelaez, ``Fast and slow light in zigzag microring resonator chains,'' Opt. Lett., vol. 34, pp. 626-628, March 2009.


\bibitem{chamorro} P. Chamorro-Posada, ``Q-enhanced racetrack microresonators,'' Opt. Commun, vol. 387, pp. 70-78, March 2017. 



\bibitem{hiremath} K.R. Hiremath, M. Hammer, R. Stoffer, L. Prkna, and J. Ctyroky, `` Analytic approach to dielectric optical bent slab waveguides,'' Opt. Quantum Electron. vol. 37, pp. 37--61, Jan. 2005.

\bibitem{kitoh} T. Kitoh, N. Takato, M. Yasu, and M. Kawachi, ``Bending loss reduction in Silica-Based Waveguides by Using Lateral Offsets,'' J. Lightwave Technol., vol.  13, pp. 555-562, Apr. 1995.



\bibitem{radiacion} P. Chamorro-Posada, ``Radiation in bent asymmetric coupled waveguides,'' Appl. Opt., vol. 58, June 2019.   


\bibitem{van} V. Van, P.P. Absil, J.V. Hryniewicz, P.-T. Ho, ``Propagation loss in single-mode GaAs-AlGaAs microring resonators measurement and model,'' J. Lightwave Technol., vol. 19, pp. 1734--1739 (2001).
  
  \bibitem{wgms3d} http://www.soundtracker.org/raw/wgms3d

  
\bibitem{krause} M. Krause, ``Finite-Difference Mode Solver for Curved Waveguides With Angled and Curved Dielectric Interfaces,'' J. Lightwave Technol., vol. 29, pp. 691--699, March 2011.

\bibitem{xia} F. Xia, L. Sekaric, and Y. A. Vlasov, ``Mode conversion losses in silicon-on-insulator photonic wire based racetrack resonators,'' Opt. Express, vol. 14, no. 9, pp. 3872--3886, 2006.


\bibitem{polarizadores} P. Chamorro-Posada, ``Ultracompact integrated polarizers using bent asymmetric coupled waveguides,'' Opt. Lett., vol.  44, pp.  2040-2043, Apr. 2019.

  \bibitem{VLC} http://www.imb-cnm.csic.es/index.php/en/clean-room/silicon-nitride-technology

 \bibitem{malitson} I. H. Malitson, ``Interspecimen comparison of the refractive index of fused silica,'' J. Opt. Soc. Am. 55, 1205-1208 (1965)


\bibitem{luke}  K. Luke, Y. Okawachi, M. R. E. Lamont, A. L. Gaeta, M. Lipson, ``Broadband mid-infrared frequency comb generation in a Si3N4 microresonator,'' Opt. Lett. 40, 4823-4826 (2015)







  
\end{thebibliography}
\end{document}